\documentclass[12pt]{article}
\pdfoutput=1 

\usepackage{cancel,slashed}
\usepackage{bbm}
\usepackage{epsf}
\usepackage{amssymb}
\usepackage{amsfonts}
\usepackage{amsmath}
\usepackage{graphicx}
\usepackage{cite}

\usepackage{latexsym}
\usepackage{color}
\usepackage{xypic}
\usepackage{cancel,slashed}
\usepackage[linktocpage]{hyperref}
\usepackage{color}
\usepackage{transparent}

\newcommand{\bmat}{\left(\begin{array}}
\newcommand{\emat}{\end{array}\right)}

\def\p{\partial}
\def\a{\alpha}

\def\b{\beta}
\def\g{\gamma}
\def\d{\delta}

\def\om{\omega}
\def\Om{\Omega}

\def\-{\hphantom{-}}

\def\s2{\frac{1}{\sqrt2}}

\def\oh{\frac{1}{2}}
\def\beq{\begin{equation}}
\def\eeq{\end{equation}}
\def\beqa{\begin{eqnarray}}
\def\eeqa{\end{eqnarray}}

\def\im{{\rm Im \,}}
\def\re{{\rm Re \,}}
\def\tr{{\rm tr \,}}

\def\Dsl{\,\raise.15ex\hbox{/}\mkern-13.5mu D} 

\def\CK {{\cal K}}
\def\CM {{\cal M}}
\def\CR {{\cal R}}
\def\CN {{\cal N}}
\def\CF {{\cal F}}

\def\CV {{\cal V}}

\def\CO {{\cal O}}

\def\re{\mbox{Re}}
\def\im{\mbox{Im}}
\def\tr{\mbox{Tr}}

\def\be{\begin{equation}}
\def\ee{\end{equation}}
\def\bea{\begin{eqnarray}}
\def\eea{\end{eqnarray}}
\def\raw{\rightarrow}
\def\Raw{\Rightarrow}

\def\bes{\begin{subequations}}
\def\ees{\end{subequations}}

\def\IZ{\mathbb{Z}}
\def\IR{\mathbb{R}}

\def\Id{{\mathbb{I}}}

\def\X{{\bf X}}
\def\oh{\frac{1}{2}}
\def\a{{\alpha}}
\def\b{{\beta}}
\def\d{{\delta}}
\def\eps{{\epsilon}}

\def\lam{{\lambda}}
\def\Om{{\Omega}}
\def\om{{\omega}}
\def\sig{{\sigma}}

\def\g{{\gamma}}

\def\p{{\partial}}

\def\sm2{{\mbox{\small 2}}}

\begin{document}
\pagestyle{plain}

\makeatletter
\@addtoreset{equation}{section}
\makeatother
\renewcommand{\theequation}{\thesection.\arabic{equation}}
\pagestyle{empty}
\rightline{IFT-UAM/CSIC-15-126}
\rightline{MPP-2015-277}
\vspace{0.5cm}
\begin{center}
\LARGE{{D6-branes and axion monodromy inflation}
\\[10mm]}
\large{Dagoberto Escobar, Aitor Landete, Fernando Marchesano, \\[2mm]}
\small{\it Instituto de F\'{\i}sica Te\'orica UAM-CSIC, Cantoblanco, 28049 Madrid, Spain \\[5mm] }
\large{and Diego Regalado  \\[2mm]}
\small{\it Max-Planck-Institut f\"ur Physik, F\"ohringer Ring 6, 80805 Munich, Germany\\[1mm]}
\small{\it Institute for Theoretical Physics and
Center for Extreme Matter and Emergent Phenomena,\\
Utrecht University, Leuvenlaan 4, 3584 CE Utrecht, The Netherlands\\[15mm]}
\small{\bf Abstract} \\[5mm]
\end{center}

\begin{center}

\begin{minipage}[h]{15.0cm} 

We develop new scenarios of large field inflation in type IIA string compactifications in which the key ingredient is a D6-brane that creates a potential for a B-field axion. The potential has the multi-branched structure typical of F-term axion monodromy models and, near its supersymmetric minima, it is described by a 4d supergravity model of chaotic inflation with a stabiliser field. The same statement applies to the D6-brane Wilson line, which can also be considered as an inflaton candidate. We analyse both cases in the context of type IIA moduli stabilisation, finding an effective potential for the inflaton system and a simple mechanism to lower the inflaton mass with respect to closed string moduli stabilised by fluxes. Finally, we compute the B-field potential for trans-Planckian field values by means of the DBI action. The effect of Planck suppressed corrections is a flattened potential which, in terms of the compactification parameters, interpolates between linear and quadratic inflation. This renders the cosmological parameters of these models compatible with current experimental bounds, with the tensor-to-scalar ratio ranging as $0.08 \lesssim r \lesssim 0.12$.

\end{minipage}
\end{center}
\newpage
\setcounter{page}{1}
\pagestyle{plain}
\renewcommand{\thefootnote}{\arabic{footnote}}
\setcounter{footnote}{0}


\tableofcontents

\newpage

\section{Introduction}
\label{s:intro}

The quest for realising models of large field inflation in supergravity and string theory has undoubtedly drawn significant attention over the past few years. Lately this pursuit has increased its activity, mainly due to the recent experimental data that can put all these models to the test \cite{Ade:2014xna,Ade:2015tva,Ade:2015lrj}. 

In the context of string theory models of inflation \cite{Baumann:2014nda,Westphal:2014ana}, one of the most promising proposals to achieve a successful model of large field inflation is to implement the idea of axion monodromy \cite{Silverstein:2008sg}. Under this principle, an axion with sub-Planckian decay constant attains to drive inflation along a trans-Planckian excursion after some effect has broken its periodicity by means of creating a non-periodic, multi-branched potential for it. The agent that creates such potential will depend on each particular scenario. For instance, one of the earliest proposals for implementing this idea involves including a brane-anti-brane system whose energy depends on the axion vev \cite{McAllister:2008hb} (see also \cite{Berg:2009tg,Palti:2014kza,Retolaza:2015sta}).

More recently, it has been realised that one may create such multi-branched potential by means of compactifications whose low energy superpotential depends on the axion. These models, dubbed F-term axion monodromy in \cite{Marchesano:2014mla}, have the advantage that they directly connect with the 4d framework in \cite{Kaloper:2008fb,Kaloper:2011jz} and that at small inflaton vevs they are described by 4d supergravity models of F-term inflation. Finally, for this class of models one may implement the machinery developed in moduli stabilisation programme \cite{Grana:2005jc,Douglas:2006es,Denef:2007pq,Quevedo:2014xia} in order to generate the axion multi-branched potential. As a consequence in the initial models of F-term axion monodromy inflation \cite{Marchesano:2014mla,Blumenhagen:2014gta,Hebecker:2014eua} the agent generating the axion potential was mostly the presence of background fluxes. The same applies to subsequent proposals along these lines \cite{Ibanez:2014kia,Hassler:2014mla,McAllister:2014mpa,Franco:2014hsa,Ibanez:2014swa,Garcia-Etxebarria:2014wla,Blumenhagen:2015xpa}.

It was however pointed out in \cite{Escobar:2015fda} that background fluxes are not the only source of F-term axion potentials suitable to build models of large field inflation. Indeed, if one considers a 4d type II compactification where a D($p+3$)-brane wraps an internal  $p$-cycle with certain a topology then a bilinear superpotential of the form \cite{Marchesano:2014iea}
\be
W_{\rm inf} \, =\, X S
\label{introWinf}
\ee
is generated. One of these two fields belongs to the open string sector and the other one to the closed string sector, and both contain an axion that may be the inflaton candidate. 

Remarkably, the superpotential (\ref{introWinf}) was proposed in \cite{Kawasaki:2000yn} as an elegant way to embed chaotic inflation in 4d supergravity. According to this proposal, extensively analysed and generalised in the supergravity literature \cite{Kallosh:2010ug,Kallosh:2010xz,Kallosh:2014vja,Nakayama:2014wpa}, the vev of the stabiliser field $S$ dynamically vanishes during inflation. Hence the axion piece of $X$ may take trans-Planckian vevs while retaining $W_{\rm inf} =0$, and therefore an F-term scalar potential bounded from below. 

The purpose of this paper is to develop and study in further detail the proposal put forward in \cite{Escobar:2015fda} to build models of large field inflation based on the above D-brane bilinear superpotential. Besides providing a UV string theory completion of the model in \cite{Kawasaki:2000yn}, the motivation for considering the setup in \cite{Escobar:2015fda} is two-fold, and directly related to the problems encountered when trying to construct models of large field inflation with only background fluxes \cite{Blumenhagen:2014nba,Hayashi:2014aua,Hebecker:2014kva}. On the one hand, in this setup the source for the inflaton F-term potential (the D-brane) is different from the source of moduli stabilisation F-term potential (the background fluxes). This allows to lower the inflaton mass with respect to that of the compactification moduli without having to fine-tune the superpotential or the point at which the compactification moduli are stabilised. As we will see, one may instead create this mass hierarchy by taking into account that the inflaton mass depends on the kinetic terms of the open string field entering (\ref{introWinf}), which are typically modified by placing the D-brane in a strongly warped region of the compactification. On the other hand, because during inflation $W_{\rm inf}$ does not change, the F-terms for these heavier compactification fields will not directly depend on the vev of the inflaton, so one expects that the backreaction issues that usually appear in supergravity models of large field inflation are alleviated, see e.g. \cite{Buchmuller:2014vda,Buchmuller:2015oma,Dudas:2015lga}.

While these ideas can be implemented both in type IIB and type IIA 4d orientifold compactifications, we will focus on the latter to develop our scenarios. In this case the source for the superpotential (\ref{introWinf}) is a D6-brane wrapping an internal three-cycle of the compactification, a setup already considered in \cite{Marchesano:2014mla}. The fields entering (\ref{introWinf}) are then the complexified D6-brane position (which contains a Wilson line) and a complexified K\"ahler modulus (which contains a B-field axion). Depending on the details of the compactification either the Wilson line or the B-field axion may be the scalar driving inflation, while the other scalar will be part of the stabiliser field. In both cases we analyse the interplay of moduli stabilisation with the inflaton sector, finding that under certain assumptions one can recover the effective potential that would arise in the supergravity model in \cite{Kawasaki:2000yn}, plus some extra terms that are remnant of the moduli stabilisation potential and that help to stabilise the inflationary trajectory. Finally, for the case of the B-field axion we are able to compute $\alpha'$ corrections to the quadratic potential that arises from $W_{\rm inf}$ by evaluating the B-field dependence of the D6-brane DBI action. Similarly to the 5-brane case in \cite{McAllister:2008hb}, we find a square-root structure that tends to a linear potential for large values of the inflaton. As a result,   depending on compactification parameters like the typical size of the internal manifold and the internal volume of the D6-brane, we find a set of models whose potentials interpolate between a linear and a quadratic behaviour. In terms of the cosmological parameters of these models, this translates into a tensor-to-scalar ratio that ranges in the values $0.08 \lesssim r \lesssim 0.12$, and a spectral index ranging as $0.962 \lesssim n_s \lesssim 0.972$, in agreement with the latest experimental bounds \cite{Ade:2015tva,Ade:2015lrj}. 

The paper is organised as follows. In section \ref{sec:axions} we review the axions that are present in type IIA orientifold vacua and how they enter the K\"ahler potential and superpotential in the presence of stabilising effects like background fluxes. In section \ref{s:lifting} we analyse the D6-brane DBI action to see how its presence can stabilise a B-field axion. We also recover the superpotential (\ref{introWinf}) in the small field limit and discuss how it is embedded in the multi-branched structure of the actual potential for the B-field and the Wilson line scalars. In section \ref{s:scenarios} we take profit of the supergravity description of F-term axion monodromy vacua to analyse the interplay between closed string moduli stabilisation and the inflaton sector, deriving an effective potential for the latter which we use to estimate the stability for the inflationary trajectory. We also discuss how to generate a hierarchy of mass scales between these two sector by means of warping. Finally, in section \ref{s:largeb} we analyse the potential at large field values for of the scenario where the B-field is the inflaton, which displays a clear flattening effect,  and we compute the slow-roll parameters of this model and compare it with the current experimental bounds. We draw our conclusions in section \ref{s:conclu}.

A substantial fraction of technical details is relegated to the appendices. Appendix \ref{ap:DBI} contains all the computations of the D6-brane DBI action necessary for the large B-field analysis. Appendix \ref{ap:4dsugra} contains the supergravity computations and identities needed in section \ref{s:scenarios} and Appendix \ref{WLtoy} contains a simple moduli stabilisation example suitable to embed the Wilson line scenario.

\section{Axions in type IIA vacua}
\label{sec:axions}

Let us consider type IIA string theory compactified in a six-dimensional manifold $\X_6$, in the presence O6-planes wrapping three-cycles of $\X_6$ and filling the remaining four spacetime dimensions. This can be achieved by first compactifying the theory on a Calabi-Yau three-fold $\CM_6$ and then modding it out by the orientifold symmetry $\CO = (-1)^{F_L} \Omega_p \CR$, where $F_L$ is the left-moving spacetime fermion number, $\Om_p$ is the world-sheet parity operator and $\CR$ is an isometric involution of $\CM_6$ acting as
\be
\CR\, J \, = \, -J \quad \quad \quad \CR\, \Om \, = \, \overline \Om 
\ee
where $J$ and $\Omega$ are respectively the K\"ahler form and the holomorphic (3,0)-form of $\CM_6$. An O6-plane will be located at the fixed point locus of $\CR$, filling as well the for non-compact dimensions $X_4$. In the absence of background fluxes, the RR tadpoles induced by the O6-plane can be cancelled by D6-branes wrapping special Lagrangian three-cycles, leading to $\CN=1$ chiral compactifications to 4d Minkowski \cite{thebook,reviewsD6}. 

The closed string moduli space of type IIA Calabi-Yau orientifolds has been thoroughly analysed in the literature, see for instance \cite{Grimm:2004ua,kt11,Grimm:2011dx}. It contains in particular $h^{1,1}_-(\CM_6)$ axions arising from the dimensional reduction of the B-field. More precisely we have $h^{1,1}_-$ complexified K\"ahler moduli $T^a$ defined by 
\be
J_c\, =\, B + ie^{\phi/2} J\, =\, T^a  \omega_a
\ee
where $\phi$ is the 10d dilaton, $J$ is computed in the Einstein frame, $l_s^{-2} \omega_a$ are harmonic representatives of $H^2_-(\CM_6, \IZ)$ and $l_s = 2\pi \sqrt{\a'}$. At the supergravity level the real parts of of each $T^a$ enjoy a continuous shift symmetry, as is manifest from the K\"ahler potential
\be
K_K \,  = \, -{\rm log} \left(\frac{i}{6} \CK_{abc} (T^a - \bar{T}^a)(T^b - \bar{T}^b)(T^c - \bar{T}^c) \right) 
\label{KK}
\ee
where $\CK_{abc} = l_s^{-6} \int_{\CM_6} \om_a \wedge \om_b \wedge \om_c \in \IZ$ are the triple intersection numbers of the compactification manifold. One can argue that perturbative $\alpha'$ corrections will not spoil such symmetry \cite{Palti:2008mg}, while exponentially suppressed corrections arising from closed string worldsheet instantons are expected to break it to a discrete subgroup.\footnote{We assume that $g_s$ corrections to this K\"ahler potential are negligible in the weak coupling regime in which we will be working.}

The shift symmetry is also broken by the presence of background fluxes. More precisely including RR background fluxes will generate a superpotential of the form
\be
l_sW_K(T)\, =\, e_0 + e_a T^a + \oh \CK_{abc} m^a T^b T^c  - \frac{1}{6} m^0 \CK_{abc} T^a T^b T^c
\label{WK}
\ee
where $(e_0, e_a, m^a, m^0)$ are integer numbers that correspond to the RR flux quanta of $(F_6, F_4, F_2, F_0)$ respectively, see \cite{Grimm:2004ua} for their precise definition and \cite{Palti:2008mg} for how $\a'$ corrections modify the value of these flux parameters. One may generalise this superpotential by adding NS three-form fluxes and metric fluxes, after which a superpotential dependence on the dilaton and complex structure moduli will appear \cite{Derendinger:2004jn,Kachru:2004jr,Villadoro:2005cu,Camara:2005dc}. Notice that adding metric fluxes will take us to the realm of non-K\"ahler orientifold compactifications, whose effective theory via Kaluza-Klein reduction has not been derived in full generality. Nevertheless, one does not expect that adding such fluxes will modify the above K\"ahler potential (up to one-loop or warping effects that we are neglecting) and in particular its shift symmetries. The same applies to the K\"ahler potential for the dilaton and complex structure moduli, which following \cite{kt11} in the absence of open string moduli can be written as\footnote{In our conventions there is an extra factor of 1/8 multiplying $\CF_{{K}{L}}$ as compared to the  expression in \cite{kt11}. This factor can be removed via a constant K\"ahler trasformation that would nontheless also affect the superpotential (\ref{WK}). In \cite{Grimm:2004ua} this factor is absorbed into the definition of $\im \, N^K$.}
\be
K_Q \, = \, -2\, {\rm log} \left(\frac{1}{16i} \CF_{{K}{L}} \left[N^{{K}} - \bar{N}^{{K}}\right]  \cdot \left[N^{{L}} - \bar{N}^{{L}}\right] \right)
\label{KQ}
\ee
Here $\CF_{KL} = \p_K\p_L \CF$ stands for the second derivative of the prepotential inherited from the $\CN=2$ unorientifolded theory. The fields $\im\, N^K$, $K = 0, \dots, h^{2,1}$ represent the dilaton and complex structure moduli, and are complexified by the periods of the RR potential $C_3$. More precisely we have that
\be
N^{K}\, =\, l_s^{-3} \int_{\CM_6} \Omega_c \wedge \b^K
\label{defNK}
\ee
with $l_s^{-3} \b^K \in H_-^3 (\CM_6, \IZ)$ is a basis of odd integer three-forms, $\Omega_c = C_3 + i \re (e^{-\phi} C \Omega)$ is the calibration for BPS three-cycles and $C \equiv e^{\oh(K_{CS}-K_{K})}$ with $K_{CS} = - {\rm log}\left( \frac{i}{l_s^6} \int \Omega \wedge \bar{\Omega}\right)$. We refer the reader to \cite{Grimm:2004ua,kt11,Grimm:2011dx} for further details and more precise definitions. 

Notice that the K\"ahler potential (\ref{KQ}) also displays a shift symmetry for those scalars arising from $C_3$, which are also axions as expected from general arguments \cite{Dimopoulos:2005ac}. In order to stabilise the fields $N^K$ it is necessary to introduce $H_3$ and metric fluxes in order to include them in the flux superpotential that generalises (\ref{WK}), to take into account the non-perturbative superpotential generated by D2-brane instantons as in \cite{Palti:2008mg}, or to project most of them out via some orbifold action as in \cite{DeWolfe:2005uu}. 

Including D6-brane moduli will modify the K\"ahler potential. Recall that given a D6-brane wrapping a smooth special Lagrangian three-cycle $\Pi_3 \in \CM_6$ its moduli space is given by a mixture of Wilson lines and geometric deformations of $\Pi_3$, and its complex dimension is given by $b_1(\Pi_3)$. More precisely, for $b_1(\Pi_3) = 1$ we define the complexified D6-brane deformation as \cite{McLean,Hitchin:1997ti,Camara:2011jg}
\be
\Phi_{\rm D6}\, =\, \frac{l_s}{\pi}\left (A - \iota_{\varphi X} J_c|_{\Pi_3} \right ) \, =\, \frac{l_s}{\pi}(\xi -\lam \varphi) \zeta \, =\,  \Phi \zeta
\label{defPhi}
\ee
with $\zeta/2\pi l_s$ the harmonic one-form generating $H^1(\Pi_3, \IZ)$, and $X$ a normal vector to $\Pi_3$ such that $e^{\phi/2} \iota_{X} J = e^{\phi/2} (X^m J_{mn}) dx^n = \zeta$, which  implies that $\iota_{X}J_c|_{\Pi_3}=\lam \zeta = (\eta + i) \zeta$ with $\eta \in \IR$. Finally, $A =  \xi \zeta$ describes the D6-brane Wilson line profile so $\xi$ has period $1/l_s$ and $\re\,\Phi$ has period $1/\pi$.\footnote{In our conventions $\int_{\pi_2} F \in 2\pi \IZ$ for every 2-cycle $\pi_2$, from where $A\sim A+l_s^{-1}\zeta$ for $\zeta/2\pi l_s\in H_1(\Pi_3,\IZ)$.} 
As argued in \cite{Blumenhagen:2006xt}, these open string fields may also enter into the non-perturbative superpotential generated by D2-brane Euclidean instantons.\footnote{This description of the D6-brane moduli space is not only valid when $\CM_6$ is a Calabi-Yau three-fold, but also for non-K\"ahler manifolds $\X_6$ arising in type IIA flux vacua \cite{Marchesano:2006ns,Koerber:2006hh}.}

Considering such open string modulus and performing a direct dimensional reduction of the D6-brane DBI action one finds that the tree-level K\"ahler potential (\ref{KQ}) is naively modified to \cite{kt11}
\be
K_Q \, = \, -2\, {\rm log} \left(\frac{1}{16i} \CF_{{K}{L}} \left[N^{{K}} - \bar{N}^{{K}} + \frac{i}{4} Q^{{K}} \Phi \bar{\Phi} \right]  \cdot \left[N^{{L}} - \bar{N}^{{L}} + \frac{i}{4} Q^{{L}} \Phi \bar{\Phi} \right] \right)
\label{KQPhi}
\ee
where $Q^K = l_s^{-3} \int_{\Pi_3} \iota_X \b^K \wedge \zeta$ and now $N^K$ depends on $\Phi$. Notice that in this expression there is no obvious shift symmetry for the open string modulus. However, from the general arguments of \cite{Kachru:2000ih} one expects the D6-brane Wilson line $A$ to only appear through its derivatives in the 4d effective action, and therefore to exhibit a shift symmetry in the K\"ahler potential. This expectation is further sustained by the fact that D6-brane Wilson lines lift to integrals of the three-form $A_3$ over three-cycles in $G_2$ compactifications of M-theory, and that such scalars are absent in the corresponding K\"ahler potential \cite{Beasley:2002db}, just like the scalars arising from $C_3$ are absent in (\ref{KQ}). Hence, following \cite{Hebecker:2012qp} one may propose a alternative K\"ahler potential combining complex structure and D6-brane moduli and which yields the same kinetic terms, namely
\be
K_Q' \, = \, -2\, {\rm log} \left(\frac{1}{16i} \CF_{{K}{L}} \left[N^{{K}} - \bar{N}^{{K}} - \frac{i}{8} Q^{{K}} (\Phi - \bar{\Phi})^2 \right]  \cdot \left[N^{{L}} - \bar{N}^{{L}} - \frac{i}{8} Q^{{L}} (\Phi -  \bar{\Phi})^2 \right] \right)
\label{KQPhi2}
\ee
in which the Wilson line shift symmetry is manifest. Absent some definite explicit computation which selects one K\"ahler potential versus the other, we will consider both possibilities throughout our discussion. As we will see, very few of our results will depend on which of the two versions of $K_Q$  is considered. 

Finally, in addition to the flux and non-perturbative superpotentials, there will be a superpotential generated by worldsheet instantons, and that may affect both the K\"ahler and open string moduli of the compactification. On the one hand we will have closed string worldsheet instantons wrapping spheres of $\CM_6$ and generating superpotential terms of the form exp $(i m_a T^a)$. These terms are suppressed by a factor exp$(-A/\a')$, with $A$ the string frame volume of a holomorphic two-cycle of $\CM_6$, so in the supergravity large volume regime they will be subleading compared to the superpotential terms discussed previously. Nevertheless, they will also contribute to the scalar potential for K\"ahler moduli and in particular one expects that they generate a periodic sinusoidal-like potential for a B-field axion. On the other hand there may also be a superpotential generated by open string worldsheet instantons, see e.g. \cite{Kachru:2000ih}. In general these will be disk instantons whose boundary lie on the non-trivial one-cycle of the D6-brane three-cycle $\Pi_3$. Such instantons will generate  superpotential terms involving the D6-brane modulus $\Phi$ and the K\"ahler moduli $T^a$. Analogously to closed string instantons, disk instantons may generate sinusoidal-like potentials for D6-brane Wilson lines.\footnote{Interestingly, such instantons may be less suppressed than their closed string counterpart in the large volume regime. Indeed, in this case $A$ will be the area of a disk, whose boundary is not determined by any volume of the compactification but rather by the position of a D6-brane in its moduli space. Therefore, even for manifolds with large two-cycles one could conceive a setup where a D6-brane has a small one-cycle and a holomorphic disk with a boundary on it, providing a non-negligible contribution to the superpotential. In the following we will not consider this possibility, although it would be interesting to engineer constructions where this could happen.}

To summarise, we have seen that there may be three different kinds of axions in type IIA vacua: B-field axions, $C_3$ RR-axions and D6-brane Wilson lines, each of them developing different superpotential terms. B-field axions develop a tree-level polynomial superpotential that may be used to generate chaotic inflation upon the inclusion of RR background fluxes, while for $C_3$ axions this can be achieved by including NS and metric fluxes/torsion in cohomology.\footnote{For D6-brane Wilson lines one may also achieve quadratic superpotentials if one introduces torsional homology in the 3-cycle wrapped by the D6-brane \cite{Marchesano:2006ns}. This case, dubbed {\it massive Wilson lines} in \cite{Marchesano:2014mla}, will not be considered here.} Schematically we have that the different pieces of superpotentials arrange as
\be
W_{\rm mod}\, =\, W_{\rm flux}(T,N) + W_{\rm D2}(N,\Phi) + W_{\rm WS}(\Phi,T)\, .
\label{Wmod}
\ee
where $W_{\rm flux}$ is the superpotential generated by the closed string fluxes threading $\CM_6$,  $W_{\rm D2}$ is the superpotential generated by Euclidean D2-brane instantons and $W_{\rm WS}$ is the correction generated by worldsheet instantons. 

Following the general philosophy of \cite{Marchesano:2014mla}, we would like to build a model of large field inflation via a superpotential involving an axion and leading to chaotic inflation. In this sense it would seem that the inflaton should be one of the fields that enter $W_{\rm flux}$. The challenge would then be to single out an axion which is much lighter than the rest of the moduli of the compactification, in order to decouple the latter from the inflationary potential. Such goal seems however quite difficult to achieve, as has been discussed in the setup of type IIB flux compactifications \cite{Blumenhagen:2014nba,Hayashi:2014aua,Hebecker:2014kva}. However, as we will discuss next there are further sources of polynomial superpotentials in type IIA vacua, which do not arise from background fluxes but rather from the presence of certain D-branes. As we will see, this will allow to develop a bilinear superpotential in which two of the above axions (namely B-field and Wilson line axions) are involved, and to build chaotic inflation scenarios for both of them.

\section{Lifting axions with D6-branes}
\label{s:lifting}

Absent background fluxes, the RR charge induced by O6-planes must be cancelled by the presence of space-time filling D-branes. 
The simplest possibility\footnote{See \cite{Font:2006na} for type IIA models which cancel tadpoles by also including coisotropic D8-branes.} is to consider $K$ stacks of D6-branes such that $N_a$ D6-branes wrap the three-cycle $\Pi_3^a$ and the following condition
\be
\sum_{a=1}^K N_a [\Pi_3^a] \, =\, 4 [\Pi_3^{O6}]
\label{tadpoleD6}
\ee
is satisfied. Here $\Pi_3^{O6}$ stands for the fixed point set of the isometric involution $\CR$, and the brackets denote the homology class of each three-cycle. By construction the whole set of D6-branes must be invariant under the orientifold action, so if $\Pi_3^b$ is not left invariant by the action of $\CR$ there must be $N_b$ D6-branes wrapping the three-cycle $\Pi_3^{b'} = \CR(\Pi_3^b)$, with the index $a$ in (\ref{tadpoleD6}) running over both stacks of branes.

The RR tadpole cancellation condition (\ref{tadpoleD6}) is sensitive to the homology class of each of the three-cycles $\Pi_3^a$ in $\CM_6$ but it is not sensitive to the topology of each three-cycle itself. In particular, a priori it tells us nothing about the number of harmonic one-forms within each three-cycle, that is about $b_1(\Pi_3^a)$. As mentioned above, such topological number indicates the number of complex open string moduli of an isolated D6-brane. More generally, it indicates the number of 4d chiral fields in the adjoint of the gauge group obtained 
from KK reduction of a stack of $N_a$ D6-branes. For that reason, when building models of particle physics, the three-cycles $\Pi_3^a$ describing the SM sector are chosen such that either $b_1(\Pi_3^a) =0$ or else the adjoint fields are projected out by some orbifold action \cite{Blumenhagen:2005tn,Forste:2008ex,Honecker:2012qr,Ecker:2014hma}, and the same is often required for the remaining D6-branes of the model. 

Nevertheless, for the purposes of this work we will consider type IIA compactifications with at least one D6-brane wrapping a three-cycle $\Pi_3^\a$ with $b_1(\Pi_3^\a) = 1$. For simplicity we will consider that such three-cycle is isolated from the rest, in the sense that it does not intersect the other three-cycles $\Pi_3^{a\neq\a}$ of the compactification, including its orientifold image. For this D6-brane to be supersymmetric it must satisfy the standard BPS conditions \cite{thebook,reviewsD6}
\bea
\label{Fterm}
J_c|_{\Pi_3^\a} - \frac{\l_s^2}{2\pi} F & = & 0 \\ 
\im\, \Omega|_{\Pi_3^\a} & = & 0
\label{Dterm}
\eea
which require that $\Pi_3^\a$ is a special Lagrangian three-cycle and that the gauge invariant field strength $\CF = B|_{\Pi_3^\a} - \frac{\l_s^2}{2\pi} F$ vanishes on it. Since $b_1(\Pi_3^\a) = 1$, $\Pi_3^a$ contains a harmonic one-form $\zeta$ and a Poincar\'e dual two-cycle $\pi_2$. It may then either happen that $\pi_2$ is homologically trivial or non-trivial in the ambient space $\CM_6$. If $\pi_2$ is trivial then any bulk closed two-form will integrate to zero over it. As a consequence the pull-back of the B-field on $\Pi_3^\a$ will be an exact one-form and so one can trivially satisfy the supersymmetry condition $\CF = 0$ by switching on the appropriate field strength $F = dA$. When moving in the moduli space of B-fields the profile for such $B|_{\Pi_3^\a} = d\beta$ will change continuously, but the condition $\CF = 0$ can always be satisfied by adjusting the profile for $A$. Hence the presence of such D6-brane does not constrain the moduli space of B-field axions.

If on the contrary $\pi_2$ is non trivial in $\CM_6$ (more precisely if $[\pi_2] \neq 0$ as an element of $H_2^-(\CM_6, \IZ)$) an obstruction to changing the B-field will appear. Indeed, in that case there is a bulk harmonic two-form $\omega$ whose integral over $\pi_2$ is non-vanishing and we may in particular assume that $l_s^{-2} \int_{\pi_2} \omega  =1$. As before, switching on a B-field of the form $B = b\, \omega$ will disturb the D6-brane BPS condition (\ref{Fterm}), but now the pull-back of the B-field no longer is an exact two-form in the cohomology of $\Pi_3^\a$, as $\omega|_{\Pi_3^\a}$ necessarily contains a harmonic piece that contributes to the integral over $\pi_2$. We may now add a field strength $F$ to cancel out the B-field pull-back, but because the harmonic piece of $F$ is quantised this is only possible whenever $b \in \IZ$. As a result, when we move along this direction in the B-field moduli space we will generate a worldvolume flux $\CF = b\, \rho$ (with $\rho$ such that $l_s^{-2} \int_{\pi_2} \rho = 1$) and supersymmetry will be broken due to the presence of the D6-brane. Finally, because on general grounds (\ref{Fterm}) can be interpreted as an F-term condition in the effective 4d theory, one expects that this effect can be understood in terms of a superpotential that lifts such B-field axion. 

The latter setup was analysed in detail in \cite{Marchesano:2014iea}, where it was found that the correct superpotential describing the above dynamics is of the form
\be
W_{\rm inf} \, =\, a\,\Phi\, T
\label{Winf}
\ee
where $a$ is a constant that will be fixed later, $\Phi$ represents the D6-brane modulus in (\ref{defPhi}), and $T$ is a combination of K\"ahler moduli defined by
\be
T\, \equiv\,l_s^{-2} \int_{\pi_2} J_c \, =\, \sum_a n_aT^a 
\label{defT}
\ee
with $n_a = l_s^{-2} \int_{\pi_2} \om_a \in \IZ$. Hence as advanced, the presence of certain D6-branes supplies yet another source of superpotential for axion fields. Since the above discussion and the derivation of (\ref{Winf}) also hold in the presence of background fluxes, (\ref{Winf}) may be directly added to the expression (\ref{Wmod}). There is however an conceptual difference between (\ref{Wmod}) and (\ref{Winf}), namely that the latter source of moduli lifting arises from a localised object. Hence in the same spirit of \cite{McAllister:2008hb} one may use warping effects to lower the masses generated from $W_{\rm inf}$ as compared to those given by $W_{\rm mod}$, as will be discussed in the next section. 

Based on the latter and some further observations, in the next section we will propose two scenarios of chaotic inflation in which the inflaton mass arises form the superpotential (\ref{Winf}). Since the supergravity description that involves $W_{\rm inf}$ is only valid at small values of the inflaton field, to perform the inflationary analysis at arbitrary field values it is necessary to derive the scalar potential microscopically and including $\a'$ corrections. This can be done for the B-field axion potential by analysing the DBI action of the D6-brane, as we do in the following.

\subsection{The D6-brane action}
\label{DBICS}

The Dirac-Born-Infeld (DBI) and Chern-Simons (CS) actions of a single D6-brane read 
\be
S_{DBI}=-\mu_6 \int d^7\xi \, e^{-\phi} \sqrt{-{\rm det}\left(P[E]-\frac{l_s^2}{2\pi} F\right)}
\label{DBID6}
\ee
and
\be
S_{CS}=\mu_6 \int P\left[C\wedge e^{-B}\right]  e^{\frac{l_s^2}{2\pi} F} 
\ee
where
\be
E=e^{\phi/2}g+B\quad  \quad \quad \mu_6=\frac{2\pi}{l_s^7}\quad\quad \quad C\,=\, C_7 + C_5 + C_3 +C_1 
\ee
and $g$ is the 10d metric in the Einstein frame. Here $P[\cdot ]$ stands for the pull-back of the 10d background into the worldvolume of the D6-brane. 

One may now consider a D6-brane on $\IR^{1,3} \times \Pi_3$, where $\Pi_3$ is a submanifold of the compact six-manifold $X_6$ with $b_1(\Pi_3) =1 $, and with a non-trivial worldvolume flux $\CF = P[B] - \frac{l_s^2}{2\pi} F$. To describe the effective theory on such D6-brane let us perform the 4d Weyl rescaling
\be\label{weylr}
g_{\mu\nu}\, \raw\, \frac{g_{\mu\nu}}{\hat{V}_{X_6}/2}
\ee
where $\hat V_{X_6} = l_s^{-6} V_{X_6}$ stands for the compactification volume of the covering space in string units. Following the  computations of Appendix \ref{ap:DBI} we arrive at a 4d effective action for the scalars 
\bea
\begin{split}\label{DBIexptext}
S_{4d}=-\int d^4x\, V_0-\frac{1 }{2}\int d^4x\,(\p_\mu \varphi\,\,\,\p_\mu\xi )\,\mathbf{M}\left (\begin{array}{c}\p^\mu\varphi\\\p^\mu\xi\end{array}\right )
\end{split}
\eea
where we have neglected terms with more than two derivatives in 4d and kept only up to quadratic terms in the open string fields $(\varphi,\,\xi)$. The first term in (\ref{DBIexptext}) corresponds to the contribution of the D6-brane to the vacuum energy of the compactification, and it amounts to
\be
V_0 \, =\,  \frac{1}{2\pi \kappa_4^4\hat V_{X_6}^2} \int_{\Pi_3} d\hat{\text{vol}}_{\Pi_3} e^{\frac{3\phi}{4}} \tilde Q \sqrt{ 1 + \frac{1}{2e^\phi} \CF_{ab}\CF^{ab}}
\ee
where $d\hat{\text{vol}}_{\Pi_3}$ stands for the volume form of $\Pi_3$ in string units, and $\kappa_4^2 = l_s^2/4\pi$. In addition $\tilde Q$ is a quadratic polynomial in $(\varphi,\,\xi)$ given in Appendix \ref{ap:DBI}. As discussed there, for Calabi-Yau and flux compactifications of interest for the scenarios in the next section, this polynomial can be replaced by the identity. Finally, we need to take into account that this vacuum energy will be partially cancelled by the presence of O6-planes in the compactification. In particular, if the D6-brane wraps a special Lagrangian three-cycle, the vacuum energy will be totally cancelled whenever $\CF=0$. We therefore arrive at the following D6-brane scalar potential 
\bea
\nonumber
V_{\rm D6} & = & \frac{g_s^{3/4}}{2\pi\kappa_4^4{\hat V_{X_6}^2}} \int_{\Pi_3} d\hat{\text{vol}}_{\Pi_3}  \sqrt{ 1 + \frac{1}{2g_s} \CF_{ab}\CF^{ab}} -  l_s^{-3} \re\, {\Omega} \\
&= &  \frac{g_s^{3/4}}{2\pi\kappa_4^4\hat V_{X_6}^2} \int_{\Pi_3} d\hat{\text{vol}}_{\Pi_3} \left( \sqrt{ 1 +  g_s^{-1} \rho^2\,  b^2 } - 1 \right)
\label{DBIpotB}
\eea
where for simplicity have considered a constant dilaton. In the second line we have set $\CF = b\, \rho$, with $b\in \IR$, $\rho$ a quantised two-form of $\Pi_3$, and $\rho^2 = \oh \rho_{ab}\rho^{ab}$ its squared norm.\footnote{The precise profile of $\rho$ will be determined by minimisation of the D6-brane potential, taking into account that because $\CF = B - \sig dA$ one can always add an arbitrary exact piece to it. In the small field limit $\rho$ will be harmonic and such that $l_s^{-2} [\rho]$ generates $H^2(\Pi_3, \IZ)$. For large B-field values one can check that it will also develop an exact component whenever $d \rho^2 \neq 0$.} Moreover, we have assumed that $\Pi_3$ is an special Lagrangian three-cycle, so that $\re \,\Omega|_{\Pi_3} =  d\text{vol}_{\Pi_3}$. In general we may consider the case where $\Pi_3$ is not a Lagrangian three-cycle, and in particular that the pull-back of the K\"ahler form on $\Pi_3$ is given by $e^{\phi/2}J|_{\Pi_3} = j\,  \rho$, with $j \in \IR$. In that case we need to take into account that
\be
d\text{vol}_{\Pi_3}\, =\, \frac{ \re\, \Omega|_{\Pi_3}}{\sqrt{(\re\, \Omega|_{\Pi_3})^2}} 
\ee
and the following identity proved in Appendix \ref{ap:DBI}
\be
1\, =\, (J|_{\Pi_3})^2 + (\re\, \Omega|_{\Pi_3})^2 + (\im\, \Omega|_{\Pi_3})^2.
\label{identity}
\ee
Using these and for simplicity imposing the D-term condition $\im\, \Omega|_{\Pi_3} \equiv 0$ one obtains
\be
V_{\rm D6} \, =\, \frac{g_s^{3/4}}{2\pi\kappa_4^4{\hat V_{X_6}^2}} l_s^{-3} \int_{\Pi_3} \re\, {\Omega} \left( \sqrt{ 1 + \frac{\rho^2}{g_s  \om^2}\, (b^2 + j^2) } - 1 \right)
\label{finalpot}
\ee
where we have denoted $\om^2 \equiv (\re\, \Omega|_{\Pi_3})^2$ in order to simplify the notation. This scalar potential directly depends on the complexified K\"ahler modulus $T$ defined in (\ref{defT}), since applying the above definitions we have that
\be
|T|^2 \,=\, b^2 + j^2\,,
\label{Tsquared}
\ee
which contains all the dependence on the B-field axion $b$. Finally, an extra dependence on the K\"ahler field saxion $j$ may appear through the factor $\rho^2/\om^2$, since the induced metric of $\Pi_3$ will in general depend on $j$. 

The last term in (\ref{DBIexptext}) contains the kinetic terms for the D6-brane fields $\varphi$ and $\xi$, which include a transverse deformation for $\Pi_3$ and a Wilson line over its non-trivial one-cycle. In terms of the definition (\ref{defPhi}) we can express the complexified D6-brane field as $\Phi=\frac{l_s}{\pi}(\xi - \eta\varphi-i\varphi)$, see Appendix \ref{ap:DBI} for more details. There it is also given the explicit expression for the kinetic term matrix {\bf M} for arbitrary values of $\CF$ and $J|_{\Pi_3}$. For instance, the entry {\bf M}$_{\xi\xi}$ is given by
\be
\frac{1}{\pi\hat V_{X_6}} \frac{1}{l_s^3} \int_{\Pi_3} d\text{vol}_{\Pi_3} e^{-\phi/4}\sqrt{W_\CF} \left( \hat g^{ab}+\frac{\CF^{ac}\CF_{c}{}^b}{g_sW_\CF} \right) \zeta_a \zeta_b, \quad \quad W_\CF = {1+\frac{1}{2g_s}\mathcal F_{ab}\mathcal F^{ab}}
\ee
where as before $\zeta/2\pi l_s$ is the quantised harmonic one-form of $\Pi_3$ and $\hat g^{ab}$ is the inverse of the induced metric. Therefore, for $\CF = 0$ the kinetic term reduces to
\be
\frac{1}{\pi\hat V_{X_6}}  \frac{1}{l_s^3} \int_{\Pi_3}  d\text{vol}_{\Pi_3}\, e^{-\phi/4} g^{ab} \zeta_a \zeta_b \, =\, \frac{1}{\pi \hat V_{X_6}}  \frac{1}{l_s^3} \int_{\Pi_3} e^{-\phi/4} \zeta \wedge *_3 \zeta
\label{kinxixi}
\ee
In addition, for $\CF = J|_{\Pi_3} = 0$ we have that ${\mathbf M}_{\varphi\xi} = - \eta\,{\mathbf M}_{\xi\xi}$ and ${\mathbf M}_{\varphi\varphi} = (1+ \eta^2)\,{\mathbf M}_{\xi\xi}$. In this limit we are therefore able to identify ${\mathbf M}$ with the kinetic term $K_{\Phi\bar\Phi}$ for the complex field $\Phi$. In fact, we can derive the same kinetic term from the K\"ahler potentials discussed in the previous section. Indeed, for this let us write either (\ref{KQPhi}) or (\ref{KQPhi2}) as $K_Q  =  -2\, {\rm log} \left(\frac{i}{4} \CF_{{K}{L}} \im\, Z^{{K}} \im\, Z^{{L}} \right)$. Then it is easy to check that
\be
K_{\Phi\bar\Phi}\, \equiv\, [\p_\Phi\p_{\bar\Phi} K_Q]_{\Phi  =0}\, =\, - \frac{1}{2} \left. \frac{\CF_{KL}Q^K\im Z^L }{\CF_{KL}\im Z^K \im Z^L} \right|_{\Phi =0}
\label{KPP}
\ee
where $K_Q$ is either given by  (\ref{KQPhi}) or (\ref{KQPhi2}). Using the fact that $e^{\phi/2} \iota_X \im\, \Omega|_{\Pi_3} = - *_3 \zeta$ and
\be
i \CF_{KL} \im Z^L \b^K\, =\, e^{-\phi/4} \im \, \Omega \quad \Raw \quad i\CF_{KL}Q^K\im Z^L \, =\, \frac{1}{l_s^3} \int_{\Pi_3} e^{-\phi/4} \iota_X \im\, \Omega \wedge \zeta
\ee
\be
i \CF_{KL}\im Z^K \im Z^L\, =\, 4 e^{-\phi/2} \hat V_{X_6}
\ee
we recover (\ref{kinxixi}) from (\ref{KPP}).\footnote{\label{foot} More precisely, we have that the 4d kinetic terms are $S_{4d}^{kin}=-\frac{1}{\kappa_4^2}\int K_{\Phi\bar\Phi}\,d\Phi\wedge *d\bar\Phi$ so in the small field limit we have $K_{\Phi\bar\Phi}=\frac{\pi}{8}{\mathbf M}_{\xi\xi}$. To connect to the notation of \cite{kt11} one should replace $\im\, Z^K \raw l^{K}$.} See \cite{kt11} for a more detailed and general discussion as well for a proof of these identities. It would be interesting to see if one could also write D6-brane field kinetic terms in the form $K_{\Phi\bar\Phi}$ from a modified K\"ahler potential that is also valid for finite values of $T$. In the following we will use this explicit expression for the kinetic terms to show that, in the same limit of small $|T|$, one can understand the scalar potential (\ref{finalpot}) as an F-term potential. 

\subsection{Superpotential description}

As pointed out above, the scalar potential (\ref{finalpot}) is non-trivial only when the pull-back two-form $J_c|_{\Pi_3}$ has a harmonic component in the homology of $\Pi_3$, and this is only possible when the two-cycle $\pi_2 \subset \Pi_3$ is non-trivial in the homology of $X_6$. Since precisely in this situation is when the superpotential (\ref{Winf}) is developed, we would expect to understand the small $|T|$ limit of (\ref{finalpot}) as an F-term induced potential. Indeed, notice that in this small field limit we can expand the square root and we obtain
\be
V_{\rm D6} \, \stackrel{|T| \ll 1}{=}\,\frac{g_s^{-1/4}}{4\pi \kappa_4^4 {\hat V_{X_6}^2}}  |T|^2  l_s^{-3} \int_{\Pi_3} \rho \wedge *_3 \rho\,,
\label{smallpot}
\ee
assuming again constant dilaton. Now we would like to compare it with the usual F-term potential in $\CN=1$ supergravity. Thus, we need to use that
\be
e^K\, =\, \frac{g_s^{-1/2}}{8\hat V_{X_6}^3}\,,
\ee
together with the inverse of the kinetic terms, which from (\ref{Kinv}) and the above read
\be
K^{\Phi\bar\Phi}|_{\Phi=0} \, =\, 8 \hat V_{X_6}\, l_s^3 \left(\int_{\Pi_3} e^{-\phi/4} \zeta \wedge *_3 \zeta  \right)^{-1}\, =\, \frac{2\hat V_{X_6} g_s^{1/4}}{\pi^2} l_s^{-3} \int_{\Pi_3} \rho \wedge *_3 \rho
\ee
where we used that $\rho$ and $*_3\zeta$ are proportional in the string frame and that $\int_{\Pi_3} \rho \wedge \zeta = 2\pi l_s^3$. Therefore, in this limit we can write (\ref{smallpot}) as
\be\label{VD6sugra}
V_{\rm D6} \, \stackrel{|T| \ll 1}{=}\, \frac{1}{\kappa_4^2} e^K K^{\Phi\bar\Phi} |\p_\Phi W_{\rm inf}|^2 
\ee
after fixing the value of $a$ introduced in \eqref{Winf} to
\be\label{valuea}
a=\frac{2\pi}{l_s}.
\ee
This means that we can understand the excess energy of the D6-brane as an F-term induced potential in an $\CN=1$ Minkowski vacuum, in the same spirit as in \cite{Martucci:2006ij}.  Notice however that the scalar potential that arises from applying the supergravity formula to (\ref{Winf}) has yet another term proportional to $|\p_T W_{\rm inf}|^2 =|a|^2 |\Phi|^2$ which will stabilise the D6-brane field, and in particular the D6-brane Wilson line. As discussed in \cite{Marchesano:2014iea} the microscopic origin of this second term is more subtle, and can only be detected by taking into account global aspects of the backreaction of the D6-branes in the model. Instead of doing so, one may provide an alternative derivation of the scalar potential for $\Phi$ based on an axion-four-form Lagrangian obtained from dimensional reduction, as we will do in the following.

\subsection{Multi-branched potential and KS Lagrangian}

In the above derivation of the scalar potential for $T$ we have implicitly ignored the fact that the DBI action does not depend on the pull-back $B|_{\Pi_3}$ but rather on the gauge invariant combination $\CF = B|_{\Pi_3} - \frac{l_s^2}{2\pi} F$, where $F$ is a two-form of $\Pi_3$ that can be decomposed as
\be
\frac{l_s^2}{2\pi}  F\, =\,\frac{l_s^2}{2\pi} dA + n_F \rho, \quad \quad n_F = \IZ
\label{decompF}
\ee
In the small B-field limit, the role of $dA$ is to remove any exact piece that $B|_{\Pi_3}$ has, so that $\CF$ is a harmonic two-form of $\Pi_3$. The role of the second component of (\ref{decompF}) is to shift the value of the B-field axion $b$ by an integer $n_F$. Taking this into account one finds that in the expressions (\ref{finalpot}) and (\ref{smallpot}) one should replace $b^2 \raw (b - n_F)^2$. Or in other words that instead of (\ref{VD6sugra}) we should have
\be
V_{\rm D6} \, \stackrel{|T| \ll 1}{=}\,\frac{1}{\kappa_4^2} e^K K^{\Phi\bar\Phi} |D_\Phi W_{\rm inf} - a\, n_F|^2
\label{multisugra}
\ee
which has its minimum at $b=n_F$. Since $n_F$ can take any possible integer value, we actually have a multi-branched potential, which recovers the periodicity of the axion moduli space broken by the superpotential. Indeed, for quantised values of the B-field axion we can go back to zero energy by changing the integral of $F$, which is interpreted as a change of potential branch. The same structure is obtained in the DBI potential (\ref{finalpot}), which contains the $\a'$ corrections to the supergravity scalar potential. 

This sort of multi-branched structure for supergravity potentials has been recently studied in \cite{Bielleman:2015ina}, where it was argued that it is generally obtained when coupling 4d four-forms to axions or polynomials thereof. The simplest possibility for such coupling is of the form
\be
\int_{X_4} - \frac{Z}{2} F'_4 \wedge *F'_4 - \oh f_{\hat \xi}^2 d\hat \xi \wedge * d\hat \xi + \sqrt{Z}f_{\hat\xi} \mu\, \hat\xi F'_4 
\label{KSL}
\ee
where $\hat \xi$ is a dimensionless axion of period one given by $\hat \xi = l_s\,\xi$ and $F'_4$ a four-form in 4d whose kinetic term is $Z$. All mass dimensions are fixed by $[\sqrt{Z}] = [f_{\hat\xi}] = [\mu]  = L^{-1}$. 

Following \cite{Kaloper:2008qs} one may express this Lagrangian in terms of a shifted four-form $\tilde{F}_4$, which we then integrate out. The resulting Lagrangian contains a scalar potential of the form
\be
V\, =\,  \frac{1}{2} \left(\sqrt{Z} c + \mu f_{\hat\xi} \hat\xi \right)^2
\label{KSpot}
\ee
where $c$ is an integration constant quantised in terms of the 4d domain wall charge as \cite{Bousso:2000xa}
\be
c\, =\, \frac{e}{Z}\, n \quad \quad n \in \IZ\,.
\ee
Finally, the discrete symmetry of this theory imposes the relation $|e|\, =\, \mu f_{\hat\xi} \sqrt{Z}$, where $f_{\hat\xi}$ is the axion decay constant, and so this allows to rewrite the potential as
\be
V\, =\, \frac{1}{2} \mu^2 f_{\hat\xi}^2 (n + \hat\xi)^2 \, =\, \oh \frac{e^2}{Z} (n + \hat\xi)^2 \,.
\label{poteZ}
\ee
Such potential has the same multi-branched structure as  (\ref{multisugra}), it was proposed in \cite{Kaloper:2008fb,Kaloper:2011jz} as a 4d description of and axion-monodromy model of inflation, and recovered from dimensional reduction of string theory compactifications in the context of F-term axion monodromy in \cite{Marchesano:2014mla}. 

Notice that in the present setup we have the same ingredients as in \cite{Marchesano:2014mla}, namely some B-field $b$ and Wilson line $\hat\xi = \pi \re\, \Phi$ axions with a superpotential generating a mass for them. Therefore one would also expect to recover a multi-branched potential whose discrete symmetry is still  preserved once that $\a'$ corrections have been taken into account, as we have already obtained for the case of $b$. For the case of the Wilson line $\hat\xi$ its scalar potential is invisible to the DBI analysis performed above, but we can nevertheless recover a Kaloper-Sorbo Lagrangian from the D6-brane Chern-Simons action. Indeed, we obtain that
\be
\frac{\mu_6 l_s^2}{2\pi} \int_{X_4 \times \Pi_3} C_5 \wedge F \, =\, \frac{1}{l_s^6} \int_{X_4} \hat\xi F'_4\cdot  \int_{\Pi_3} \zeta \wedge \omega\, =\, \frac{2\pi}{l_s^3} \int_{X_4} \hat\xi F'_4
\label{KSWL}
\ee
where $F'_4 = d C_3'$ and we have decomposed the RR potential $C_5$ and D6-brane gauge potential $A$ as
\be
A\, =\,  l_s^{-1}\hat\xi \zeta \qquad \quad
C_5\, =\, C_3' \wedge \omega \qquad \omega\, =\, n_a \omega_a
\ee
and used that $\int_{\Pi_3} \zeta \wedge \omega = \int_{\Pi_3} \zeta \wedge \rho\, =\, 2\pi l_s^3$. Finally, a term of the form $-\oh \int_{X_4} Z  F'_4 \wedge * F'_4$ will arise from the dimensional reduction of the 10d kinetic term $\int (dC_5)^2$ in the 10d type IIA supergravity Lagrangian. We then recover the full Kaloper-Sorbo Lagrangian (\ref{KSL}), with
\be
Z\, =\, \frac{1}{4\kappa_4^2} g_s^{1/2}\hat V_{X_6}^3  K_{T\bar{T}}\, =\, \frac{1}{32\kappa_4^2} e^{-K} K_{T\bar{T}}
\ee
where we have used (\ref{derivKK}). Using this expression for $Z$ and (\ref{KSWL}) we deduce that the scalar potential felt by the Wilson line at small field values is 
\be
V \, = \,  \frac{1}{\kappa_4^2} e^{K} K^{T\bar{T}} \frac{a^2}{\pi^2} \left (\pi \re\, \Phi  - n \right)^2 \quad \raw \quad V \, = \,  \frac{1}{\kappa_4^2} e^{K} K^{T\bar{T}} |D_T W_{\rm inf} - l_s^{-1} 2 n |^2
\label{potWL}
\ee
with $a$ again given by \eqref{valuea}, providing an independent derivation of its value. We have extended the potential to include the saxion dependence, which can be included directly or by applying the approach in \cite{Dudas:2014pva}. Here $n$ labels each of the branches of the potential, and the $n=0$ branch is directly described by the F-term generated potential applied to (\ref{Winf}).  As usual, transition between these branches is possible via domain wall nucleations. Since the 4d three-form that these domain walls couple to arises from the dimensional reduction of the RR potential $C_5$, these domain walls must correspond to D4-branes wrapping the non-trivial two-cycle $\pi_2$ of the D6-brane that is also non-trivial in the bulk. From the microscopic viewpoint such domain walls shift the value of the internal RR flux $F_4 = dC_3$ along the four-form of $X_6$ Poincar\'e dual to $\pi_2$. As a result, in the system at hand an internal large gauge transformation on the D6-brane implies a discrete shift in the Wilson line $\pi \re\, \Phi \raw \pi \re\, \Phi + k$ and a compensating discrete shift of $F_4$ in the Poincar\'e dual of $[\pi_2]$. We will come back to such discrete invariance at the end of the next section, where we will see how imposing it at the level of the superpotential again fixes the value for $a$ to be \eqref{valuea}.

\section{Two type IIA scenarios of large field inflation}
\label{s:scenarios}

In the previous section we have analysed the scalar potential that is developed for the complex fields $\Phi$ and $T$ in the presence of a D6-brane with specific topology. In general, this will be part of a full scalar potential that should also stabilise all the other moduli. In the following we would like to discuss the interplay between the inflationary potential and the potential fixing the compactification moduli.  Our strategy will be to consider the regime where the fields $\Phi$ and $T$ have small vevs, and so the inflationary potential can be described in terms of the superpotential (\ref{Winf}). The full scalar potential is then specified by the supergravity F-term potential derived from
\be
W\, =\, W_{\rm mod}  + W_{\rm inf} 
\label{Wtot}
\ee
with $W_{\rm mod}$ given by (\ref{Wmod}) and $W_{\rm inf}$ by (\ref{Winf}). Such supergravity framework allows to see if a hierarchy of mass scales can be obtained between the inflaton candidate and all the other moduli, and to what extent taking the inflaton away from its minimum affects the stabilisation of heavier scalars. 

If one succeeds in decoupling the inflaton sector from the rest of the compactification moduli, then a natural question is whether one can recover a 4d $\CN=1$ supergravity model of chaotic inflation like those in \cite{Kawasaki:2000yn,Kallosh:2010ug,Kallosh:2010xz,Kallosh:2014vja,Nakayama:2014wpa}. Such 4d models are based on a bilinear superpotential like (\ref{Winf}), as well as on a particular sort of K\"ahler potentials that allow to give masses above the Hubble scale to all the scalars in $\Phi$ and $T$ except the inflaton. 
We shall analyse such questions in the context of type IIA compactifications with the superpotential (\ref{Wtot}), distinguishing two different kinds of scenarios. In the first scenario the inflaton candidate is the $B$-field axion within $T$, while in the second one it is the D6-brane Wilson line that lies within $\Phi$. 

\subsection{Inflating with the B-field}
\label{Bfield}

Let us first consider the case in which the inflaton candidate is the B-field axion $\re T$
\be
b\, =\, l_s^{-2} \int_{\pi_2} B\, =\, \sum_a n_a b^a
\label{bfield}
\ee
similarly to the setup explored in \cite{Escobar:2015fda}. As in there, we can single out this B-field axion from the rest by assuming that $T$ is the only combination of K\"ahler moduli that does not appear in $W_{\rm mod}$. Furthermore, we assume that the typical mass that closed string moduli acquire from $W_{\rm mod}$ is well above the Hubble scale, while those only included in $W_{\rm inf}$ acquire a smaller mass. Under such assumptions (which we will justify later on) one may integrate out all the massive moduli and be left with an effective theory for the complex fields $T$ and $\Phi$, whose dynamics will be dictated by an effective potential $V^{\rm eff}(T,\bar{T},\Phi,\bar{\Phi})$ obtained after freezing all the other moduli.\footnote{Up to backreaction effects on such frozen moduli which can be analysed following \cite{Buchmuller:2014vda}. We postpone the precise analysis of these effects  for future work, as well as the discussion of scenarios with non-vanishing F-terms, in which backreaction effects can be particularly important \cite{Buchmuller:2015oma,Dudas:2015lga}.} To this potential there should correspond an effective K\"ahler and superpotentials $K^{\rm eff}(T,\bar{T},\Phi,\bar{\Phi})$ and $W^{\rm eff}(T,\Phi)$, and whenever $\p_T W_{\rm mod} = \p_\Phi W_{\rm mod} = 0$ one would expect that the latter is of the form $W^{\rm eff} = W_{\rm inf} + W_0$. Finally, if we impose that $|W_0|$ vanishes or it is very small, the effective supergravity model falls into the category considered in \cite{Kawasaki:2000yn,Kallosh:2010ug,Kallosh:2010xz,Kallosh:2014vja,Nakayama:2014wpa}, with the field $T$ containing the inflaton and $\Phi$ being a stabiliser field. Therefore, one may take profit from the results of these references as a guideline to evaluate whether our setup may yield a successful inflationary model. 

Let us in particular consider the analysis of \cite{Kallosh:2010xz} for general K\"ahler potentials. There it is shown that if $K^{\rm eff}(T,\bar{T},\Phi,\bar{\Phi})$ is invariant under the following transformations
\bes
\begin{align}
T &\, \raw\,  \bar{T}\\
T &\, \raw \, T + c, \quad c \in \IR \\
\Phi &\, \raw \, -\Phi
\end{align}
\label{symK}
\ees
then the supergravity scalar potential $V^{\rm eff}(T,\bar{T},\Phi,\bar{\Phi})$ has an extremum at the inflationary trajectory
\be
\Phi\, =\, \im\, T\, =\, 0 
\label{trajectory}
\ee
with respect to $\Phi$, $\bar{\Phi}$ and $\im\, T$. In our setup, we may directly analyse these symmetries in the full type IIA K\"ahler potential $K = K_K + K_Q$, assuming that if present in $K$ they will also be there in $K^{\rm eff}$. On the one hand, it is then easy to check that the last two conditions in (\ref{symK}) are automatically satisfied. On the other hand, the first one is only satisfied if the intersection numbers $\CK_{abc}$ in (\ref{KK}) are chosen so that $K_K$ only depends on $(T-\bar{T})^2$, something that we will impose in the following. 

Besides checking that (\ref{trajectory}) is an extremum of $V^{\rm eff}(T,\bar{T},\Phi,\bar{\Phi})$ one should also verify that the orthogonal directions are  non-tachyonic and in particular whether the masses of the fields $\im \, T$ and $\Phi$ are above the Hubble scale. Following \cite{Kallosh:2010xz} one may do so by analysing the quartic derivatives of the effective K\"ahler potential, finding some stability bounds for the inflationary trajectory. Rather than doing so, we will carry the analysis of such stability bounds directly in terms of the effective potential $V^{\rm eff}(T,\bar{T},\Phi,\bar{\Phi})$ of this scenario, which we now turn to analyse.

\subsubsection*{Inflaton potential from a two-step approach}

To make this scenario more precise let us describe our scalar potential via a two-step approach. In the first step we consider a type IIA flux compactification with no D6-brane. The dynamics of the closed string moduli are given by the superpotential $W_{\rm mod}$ as in (\ref{Wmod}) but with $\Phi=0$ and by a K\"ahler potential which is the sum of (\ref{KK}) and (\ref{KQ}). As above, we assume that $W_{\rm mod}$ does not depend on $T$ and that $K_K$ depends on it via $(\im\, T)^2$. Finally, we assume that we can find a vacuum where all the F-terms for $T^a$ and $N^K$ vanish and with a very small or vanishing value $|W_{\rm mod}^0|$ for $|W_{\rm mod}|$ at the locus where the closed string moduli are stabilised, as in the first step of \cite{kklt}. Notice that because $\p_TW_{\rm mod} = 0$ the B-field component $\re\, T$ given by (\ref{bfield}) is a flat direction of the scalar potential. As a second consequence we have that
\be
D_T W_{\rm mod}\, =\, K_T W_{\rm mod}\, \propto\, (\im\, T) W_{\rm mod}
\ee
and so the vacuum will be reached at $\im\, T =0$. By the results of \cite{Conlon:2006tq} the direction $\im\, T$ will have a squared mass of the order $e^K |W_{\rm mod}^0|^2$, and hence very small by our assumption of small $|W_{\rm mod}^0|$.\footnote{Following \cite{Conlon:2006tq}, $\im\, T$ will be in fact either massless or tachyonic, although not implying any instability for vacuum. This point will not be a difficulty in our setup, because due to the ingredients of the second step of our construction $\im\, T$ will gain a large positive mass.} All the closed string moduli besides $T$ are assumed to have a mass above the Hubble scale due to the scalar potential derived from $W_{\rm mod}$. 

In the second step we add the D6-brane that generates the superpotential $W_{\rm inf}$. This not only shifts the superpotential to (\ref{Wtot}) but it also restores the $\Phi$ dependence of $W_{\rm mod}$. Finally, it modifies $K_Q$ to either (\ref{KQPhi}) or (\ref{KQPhi2}). All these changes will alter the expression for the scalar potential, which we can analyse around the trajectory (\ref{trajectory}). In particular, the F-terms for the complex structure moduli now read
\be
D_{N^K} W \, =\, D_{N^K} W_{\rm mod} + K_{N^K} W_{\rm inf}
\label{FNK}
\ee
with $ [D_{N^K} W_{\rm mod}]_{\Phi=0} =0$ from the first step. For the K\"ahler moduli other than $T$ we have
\be
D_{T^\a} W \, =\, D_{T^\a} W_{\rm mod} + K_{T^\a} W_{\rm inf}\, =\, K_{T^\a} W_{\rm inf} + \dots
\label{DTaW}
\ee
where in the dots contain terms beyond linear order in $\im\, T$, $\Phi$ or $\bar \Phi$. We may now plug these expressions into the 4d supergravity potential
\be
V\, =\, \kappa_4^{-2} e^K \left(K^{\a\bar{\b}}D_\a W D_{\bar{\b}} \bar{W} - 3|W|^2  \right) \quad \quad \quad \a, \b \, =\, N^K, T^a, \Phi
\label{pot}
\ee
in order to derive an effective potential $V(T,\bar{T},\Phi,\bar{\Phi})$ around the locus $\im\, T = \Phi = \bar\Phi=0$. Such computation is done in  Appendix \ref{ap:4dsugra} up to terms of quadratic order in $\im\, T$, $\Phi$, $\bar\Phi$. Because in this setup the dependence of $W_{\rm mod}$ on $\Phi$ comes through worldsheet and D-brane instantons one may consider it negligible. The result then is
\be
V \, = \, \kappa_4^{-2} e^K\left( K^{\Phi\bar\Phi} \left| \p_\Phi W_{\rm inf} \right|^2 + (K^{T\bar T}  + 4 (\re\, T)^2) |\p_TW_{\rm inf}|^2 \right)  + \CO(|W_{\rm mod}^0|)
\label{Vfinal}
\ee
where we also have neglected terms of order $|W_{\rm mod}^0|$ by using the assumption that it is a small quantity. As discussed in Appendix \ref{ap:4dsugra} the inflationary trajectory
\be
\text{Traj} = \left\lbrace \text{Re{\it T}} \neq 0  ,\,  \text{Im{\it T}} = 0 ,\,   \Phi = 0  \right\rbrace
\label{infTB}
\ee
is stable in the sense that it is a minimum of the non-inflationary directions. Indeed we have that
\be
\p_{\text{Im}\, T} V|_{\text{Traj}}\, =\, \p_{\Phi} V|_{\text{Traj}}\, =\, \p_{\bar\Phi} V|_{\text{Traj}} \, =\, 0
\ee
and that the masses for the canonically normalised saxion and stabiliser field are given by
\be
m_{\rm saxion}^{2}\mid_{\rm Traj}\, \simeq\, 6 H^2 \quad \quad \quad m_{\rm stab}^{2}\mid_{\rm Traj}\, \simeq \, 15H^2 
\label{massesB}
\ee
where $H$ is the Hubble scale, and where the second mass actually depends on the point where complex structure moduli are stabilised. Along the trajectory we find that
\be
V|_{\text{Traj}}\, =\, \frac{e^K}{2\kappa_4^4} \frac{K^{\Phi\bar\Phi}}{K_{T\bar{T}}} |a|^2\, \phi_{b}^2
\label{VTB}
\ee
where $\phi_{b} = \kappa_4 \sqrt{2K_{T\bar{T}}} \, b$ is the canonically normalised inflaton. This quadratic potential matches the one obtained from (\ref{VD6sugra}) for small values of the field $T$. As discussed in the last section for large values of the inflaton one needs to replace it by the expression (\ref{finalpot}) obtained directly form the DBI action, and which will be analysed in the section \ref{s:largeb}.

Finally, although one can make $W_{\rm flux}$ independent of $T$ upon an appropriate choice of flux quanta, the dependence of $W_{\rm mod}$ in $T$ through $W_{\rm WS}$ will always be there. As discussed in section \ref{sec:axions} such contribution is much smaller than the terms in $W_{\rm flux}$ and $W_{\rm inf}$ in the large volume regime which we are considering. Nevertheless one could expect that it gives rise to a small sinusoidal contribution superimposed over the DBI potential, similar to the effect discussed in \cite{Flauger:2009ab}.

\subsection{Inflating with a Wilson line}\label{WLine}

In this second scenario the inflaton is identified with the Wilson line $\xi$ within the D6-brane field $\Phi$, as defined in eq.(\ref{defPhi}). A shift symmetry for such Wilson line will only be realised at the level of the K\"ahler potential if it is described by $K = K_K + K_Q'$, with $K_Q'$ given by (\ref{KQPhi2}). In the following we will assume this to be the case and analyse the viability of the corresponding inflationary trajectory which now is
\be
\text{Traj} = \left\lbrace \re\, \Phi \neq 0  \text{ ,  } \im\, \Phi = 0 \text{ ,  }  T = 0  \right\rbrace
\ee
and with all the remaining closed string moduli $\{N^K, T^\a\}$ stabilised at some particular value by $W_{\rm mod}$. We may also apply the above two step approach to obtain an effective scalar potential for $\{\Phi, T\}$ around this trajectory. In fact, since now the inflaton is an open string field this two step approach is particularly convenient. Indeed, for the first step one may apply any of the schemes for type IIA closed string moduli stabilisation developed in the literature, see for instance \cite{DeWolfe:2005uu,Camara:2005dc,Palti:2008mg}. The closed string moduli will then be stabilised at some locus which should be compatible with $T=0$, otherwise the D6-brane which is introduced in the second step cannot be BPS.\footnote{It would be interesting to explore if such setup could lead to a vacuum energy uplift mechanism.} As we illustrate in Appendix \ref{WLtoy} this sort of condition is however easy to satisfy in concrete examples by appropriate choices of background fluxes, and then one recovers a superpotential of the form
\be
W_{\rm mod} \, =\, W_1 + W_2 T^2 + \dots
\ee
where $\p_T W_1 = \p_T W_2 = 0$ and the dots contain higher polynomials in $T$. 

In the second step one adds the D6-brane described in section \ref{s:lifting} and modifies the superpotential and K\"ahler potential. Because $\Phi$ only enters in $W_{\rm mod}$ via non-perturbative effects, one may consider this dependence to be negligible\footnote{More precisely one may consider that there is a sinusoidal potential for the Wilson lines that arises from the disk worldsheet instantons that contribute to $W_{\rm WS}$ in (\ref{Wmod}), small compared to the potential generated by $W_{\rm inf}$. As in \cite{Flauger:2009ab} one may treat such contribution as a superimposed modulation over the dominant potential piece.} and therefore the only agent breaking the shift symmetry for the Wilson line will be the shift in the superpotential given by $W_{\rm inf}$ or equivalently the coupling to a four-form described by (\ref{KSWL}). 

As discussed in Appendix \ref{ap:4dsugra}, applying this approach one obtains an effective potential $V^{\rm eff}(T,\bar{T},\Phi,\bar{\Phi})$ of the form
\be
V  =  \kappa_4^{-2} e^K\left( K^{\Phi\bar\Phi} \left| \p_\Phi W_{\rm inf} \right|^2 + K^{T\bar T} |\p_TW_{\rm inf} + 2TW_2^0|^2  + 4 |a|^2 (\re\, T)^2 (\re\, \Phi)^2 \right) + \CO(W_{\rm mod}^0)
\label{VfinalWL}
\ee
where $W_2^0$ is the value of $W_2$ at the point where the closed string moduli are fixed. Notice that this quantity may or may not be small depending on the specifics of closed string moduli stabilisation. If one considers it either negligible or vanishing the stability constraints for the inflationary trajectory are simplified, and we obtain 
\be
\p_{\text{Im}\, T} V|_{\text{Traj}}\, =\, \p_{\Phi} V|_{\text{Traj}}\, =\, \p_{\bar\Phi} V|_{\text{Traj}} \, =\, 0
\ee
as well as $m_{\rm saxion}^{2}\mid_{\rm Traj}\, \simeq\, 6 H^2$ and a higher mass for the stabiliser field. Finally, the inflaton potential along the above trajectory is given by
\be
V|_{\text{Traj}}\, =\, \frac{e^K}{2\kappa_4^4} \frac{K^{T\bar{T}}}{K_{\Phi\bar\Phi}} |a|^2\, \theta^2
\label{VTWL}
\ee
where $\theta$ is the canonically normalised Wilson line. This reproduces the quadratic potential obtained in the previous section either via supergravity or axion-four-form Lagrangian techniques.

Compared to the case of the B-field, a technical disadvantage of this scenario is that it is not known how to compute the Planck suppressed corrections that may modify the scalar potential for large values of the inflaton. This is because the potential that the Wilson line suffers is due to backreaction of the D6-brane into the supergravity background, and so the D6-brane action is insensitive to it \cite{Marchesano:2014iea}. Hence, even if like in \cite{Ibanez:2014swa} the inflaton is an open string field, in order to find the scalar potential for large inflaton values would imply computing the relevant $\alpha'$ corrections to the supergravity Lagrangian, which is beyond the scope of this paper. Notice however that because the potential arises from an axion-four-form Lagrangian we know that these corrections cannot be arbitrary, and that the corrected potential and kinetic terms should be expressed as powers of the initial potential itself \cite{Kaloper:2008fb,Kaloper:2011jz}. It however remains to be seen whether such corrections will lead to a flattening of the scalar potential for large values of the inflaton field and allow this scenario to be compatible with experimental data. We leave such question for future work, and in the meanwhile analyse the large field inflationary dynamics only for the B-field scenario. As we will see in the next section, in that case the flattening effect are indeed such that experimental data can be reproduced.

\subsection{Generating mass hierarchies}

One of the key assumptions of this section is the fact that all scalar fields beyond the inflaton and the stabiliser field gain a mass via $W_{\rm mod}$ which is much higher than the Hubble scale, so that we can neglect their dynamics during inflation up to a good approximation (see e.g. \cite{Buchmuller:2014vda}). In particular one would like that all those heavy closed string moduli gain a mass of at least one order of magnitude above the Hubble scale at the supersymmetric vacuum and two above the inflaton mass. In the supergravity models of chaotic inflation \cite{Kawasaki:2000yn,Kallosh:2010ug,Kallosh:2010xz,Kallosh:2014vja,Nakayama:2014wpa} related to our scenarios this can in principle be done by tuning the parameter $a$ in the inflationary superpotential (\ref{Winf}) to a small value, which allows to have an inflaton parametrically lighter than any field entering $W_{\rm mod}$. In the string constructions that we are considering this is however not possible, for reasons that we now explain. 

\subsubsection*{Why $a$ is not small}

For simplicity let us consider a type IIA compactification where the dependence of $W_{\rm mod}$ on K\"ahler moduli is contained in (\ref{WK}). Let us then add the superpotential term (\ref{Winf}) that we can write as
\be
W_{\rm inf} \, =\, a \Phi\, n_a T^a
\label{Winfa}
\ee
with $n_a\in \IZ$ as defined below (\ref{defT}). Notice that the full superpotential then satisfies
\be
W\, \supset\, W_K(e_0, e_a, m^a, m^0) + W_{\rm inf}\, =\, W_K(e_0, e_a + a l_s \Phi n_a, m^a, m^0)
\ee
or in other words, that adding $W_{\rm inf}$ can be absorbed into a redefinition of the flux superpotential integer parameter $e_a$. As a consequence we have that the superpotential is invariant under the simultaneous shift
\be
\pi \re\, \Phi\, \raw\, \pi \re\, \Phi + \frac{2\pi k}{al_s} \quad \quad \quad e_a\, \raw\, e_a - 2k n_a
\label{LGTshift}
\ee
where $k \in \IZ$ so that $e_a$ is shifted by an even integer number and flux quantisation around O-planes is left unaffected \cite{Frey:2002hf}.\footnote{We are assuming that g.c.d.$(n_a)=1$, which is typically the case for irreducible two-cycles like $\pi_2$.} This discrete shift symmetry is reminiscent of the one encountered in the branched-potential (\ref{potWL}), with now the branches being labeled by the RR four-form quanta $e_a$. Notice that this makes precise the intuitive picture developed below eq.(\ref{potWL}), where it was concluded that an integer shift of the Wilson line $\pi \re\, \Phi \raw \pi \re\, \Phi + k$ must be compensated by a corresponding shift in the RR four-form flux, and more precisely along the Poincar\'e dual of the two-cycle $\pi_2$ within the D6-brane, which corresponds to the shift $e_a \raw e_a - 2k n_a$ described above. Because this discrete Wilson line shift is a large gauge transformation, the invariance must not only be manifest at the level of the scalar potential, but also at the level of the superpotential, and this is why we can detect it via the above reasoning. Finally, the Wilson line shift in (\ref{LGTshift}) corresponds to an integer period of the Wilson line only if
\be\label{valuea2}
a=\frac{2\pi}{l_s}
\ee
as obtained independently via the expressions (\ref{VD6sugra}) and (\ref{potWL}). We however now see that the fact that $a$ is comparable to the other coefficients in the flux superpotential is not an accident of the model, but that instead is related to the discrete symmetry underlying the multi-branched potential, the same one that it is invoked in \cite{Kaloper:2008fb,Kaloper:2011jz} and in F-term axion monodromy models \cite{Marchesano:2014mla} in order to protect the scalar potential against dangerous transplanckian corrections. 

\subsubsection*{Generating hierarchies via warping}

Since the coefficient in $W_{\rm inf}$ cannot be made small, in order to generate hierarchies with respect to the closed string moduli in $W_{\rm mod}$ we are in principle left with two options.
\begin{itemize}

\item[{\it i)}] Make the coefficients in $W_{\rm mod}$ large. 

For instance, one may scale the flux quanta in (\ref{WK}) by a large integer number, in the spirit of \cite{Blumenhagen:2014nba,Blumenhagen:2015kja}. However, as this fluxes contribute to the RR tadpoles of the compactification there will typically be an upper bound for them, and so we cannot use this strategy to make these fields parametrically heavy. 

\item[{\it ii)}] Create hierarchies via the kinetic terms. 

Notice that in both of the scenarios described above the physical inflaton mass is suppressed by the open string kinetic term $K_{\Phi\bar\Phi}$, as can be seen from (\ref{VTB}) and (\ref{VTWL}). Hence, if we construct a setup in which 
\be
K_{\Phi\bar\Phi}\, \gg\, K_{\a\bar\b} \quad \quad \a, \b \, =\, N^K, T^a
\label{hierarchy}
\ee
then we will typically generate a hierarchy of masses between the inflaton sector and the fields in $W_{\rm mod}$. 

\end{itemize}

Looking at eq.(\ref{KPP}) and comparing to the kinetic terms for the closed string moduli, we see that (\ref{hierarchy}) will be easily satisfied with respect to the complex structure moduli in the limit of large complex structure. The hierarchy is not so clear with respect to the K\"ahler moduli, and in general the answer will depend on the value at which closed string moduli are stabilised.

However, taking into account that $\Phi$ is a field localised at the D6-brane worldvolume, one may consider using warping effects in order to generate a hierarchy with the closed string kinetic terms. Indeed, let us consider a type IIA flux compactification with Ansatz 
\be
ds^2\, =\, Z^{-1/2} g_{\mu\nu}^{\rm 4d} dx^\mu dx^\nu + ds^2_{X_6}
\ee
where the warp factor $Z$ only depends on the internal coordinates of $X_6$. Such backgrounds may develop regions of strong warping, like those analysed in \cite{Franco:2014hsa}, where $Z \gg 1$. If we now place the D6-brane generating the superpotential $W_{\rm inf}$ in such region, the kinetic terms for the D6-brane field $\Phi$ will be enhanced with respect to those of the closed string moduli, since the latter come from bulk integrals that are typically insensitive to warping effects. Following similar computations to those in  \cite{Marchesano:2008rg}, in simple cases one obtains an enhancement for $K_{\Phi\bar\Phi}$ which can be encoded in the rescaling of the form
\be
Q^K\, \raw \, Z_{\rm D6}^p Q^K
\label{warpZ}
\ee
where $Z_{\rm D6}$ stands for the approximate value of the warp factor at the region where the D6-brane is located, and the value of the parameter $p \in [0,1]$ depends on how the warping enters $ds_{X_6}^2$ and on the specific D6-brane embedding.\footnote{In terms of a mirror D7-brane without worldvolume fluxes, the case $p=1$ corresponds to a position modulus and the case $p=0$ to a Wilson line modulus \cite{Marchesano:2008rg}.} In any case this enhancement via warping will contribute to increase the value of the open string kinetic terms, hence decreasing the mass of the inflaton system with respect to those moduli affected by $W_{\rm mod}$. 

This effect of warping that lowers the inflaton mass can be understood intuitively in the scenario of section \ref{Bfield} where the inflaton is the B-field. Indeed, in that case the inflaton potential is generated because the pull-back of the B-field induces D4-brane  charge and tension on the worldvolume of the D6-brane, and this breaks supersymmetry. Placing the D6-brane in a region of strong warping will warp down such induced tension, flattening the potential and lowering the inflaton mass. In this sense, this mechanism for lowering the inflaton mass is analogous to the one used in \cite{McAllister:2008hb}, with our D6-brane replaced by a NS5-brane and the induced D4-brane tension with that of a D3-brane. It is however important to notice two differences with the setup in \cite{McAllister:2008hb}. First in our case the induced charge is non-conserved (simply because in generic compactifications there are no non-torsional one-cycles that a D4-brane can wrap) hence no anti-brane is needed and the caveats raised in \cite{Conlon:2011qp} do not apply. Second, as usual in models of F-term axion monodromy the system is supersymmetric at the minimum of the potential \cite{Marchesano:2014mla}, and in fact admits an effective supergravity description in the small field regime which we have worked out in the previous section. As a result in this regime the effect of warping should be understood in terms of 4d supergravity quantities. As we have seen above the coefficient in the superpotential $W_{\rm inf}$ are fixed by the discrete symmetry underlying the system, and therefore the only quantity that the warping can affect is the K\"ahler potential and more precisely the open string kinetic terms.

\subsubsection*{Scale dependence of the model}

In order to illustrate the above discussion let us see how the kinetic terms and masses for the inflaton system and the moduli in $W_{\rm mod}$ depend on the scales of the compactification. As usual the relation between the 4d Planck mass and the string scale is given by
\be
M_{\rm pl}^2\, =\, \frac{2\pi \hat{V}_{X_6}^E}{l_s^2}
\ee
where $\hat{V}_{X_6}^E$ stands for the the compactification volume in string units and in the Einstein frame.\footnote{This quantity is simply denoted by $\hat{V}_{X_6}$ in the rest of the paper, but here we make the superscript explicit in order to distinguish it from the volume measured in the string frame.} After performing the 4d Weyl rescaling
\be
g_{\mu\nu}\, \raw\, \frac{g_{\mu\nu}}{\hat{V}_{X_6}^E/2}
\ee
made in \cite{kt11} the compactification volume dependence in $M_{\rm pl}^2$ disappears and is encoded in the 4d metric. Therefore, in order to measure mass scales in Planck units we need to compare write them in terms of the mass scale $\kappa_4^{-1} = \sqrt{4\pi} l_s^{-1}$ that has appeared in several instances in the previous sections. 

To evaluate the typical value of the kinetic terms let us express the typical lengths of the compactification and of the D6-brane internal worldvolume as
\be
\hat{L}_{X_6}\, =\, \left(\hat{V}_{X_6}^E\right)^{1/6} \quad \quad \quad \hat{L}_{\Pi_3}\, =\, \left(\hat{V}_{\Pi_3}^E\right)^{1/3}
\ee
respectively. Then it is easy to see that the K\"ahler metrics for the open string and K\"ahler moduli at the minimum of the potential scale as
\be
{K}_{\Phi\bar{\Phi}} \, \sim \,  \frac{\pi^2}{2} Z^p_{\rm D6} g_s^{-1/4} \frac{\hat{L}_{\Pi_3}}{\hat{L}_{X_6}^{6}}
\ee
\be
 {K}_{T\bar{T}} \, \sim \,  \frac{1}{2} g_s^{-1} \hat{L}_{X_6}^{-4} 
\ee
respectively. At this point the inflaton potential is correctly described by 4d supergravity and so we can extract the inflaton mass for our two scenarios from either eq.(\ref{VTB}) or eq.(\ref{VTWL}). In both cases we find that the inflaton mass is given by
\be
\kappa_4^2\, m_{\rm inf}^2\, =\, e^K \left(K_{\Phi\bar\Phi}K_{T\bar{T}}\right)^{-1} |\tilde{a}|^2\, \sim \, \frac{1}{2\pi} \frac{g_s^{3/4} Z^{-p}_{\rm D6}}{\hat L_{X_6}^{8} \hat{L}_{\Pi_3}}
\ee
where $\tilde{a} = a \kappa_4$. On the other hand the typical mass of a K\"ahler modulus that appears in (\ref{WK}) will be
\be
\kappa_4^2 m_{T^\a}^2\, =\, e^K \left( K_{T\bar T} \right)^{-2} \frac{(2n)^2}{4\pi}\,\sim \,  \frac{n^2}{2\pi} \frac{g_s^{3/2}}{\hat L_{X_6}^{10}}
\ee
where $2n \in 2\IZ$ is the relevant quantum of RR flux. The quotient of both masses is then
\be
\frac{m_{T^\a}^2}{m_{\rm inf}^2}\, 
\sim\, n^2 Z^p_{\rm D6} g_s^{3/4} \frac{\hat{L}_{\Pi_3}}{\hat L_{X_6}^2}\, .
\label{ratioTinf}
\ee
%

In order to see if this dependence of the compactification scales can give an appropriate hierarchy of scales let us consider the following values
\be
\hat V_{X_6}^{\rm st} \, \sim\, 10^3 \quad \quad \hat V_{\Pi_3}^{\rm st} \, \sim\, 10 \quad \quad g_s^2\, \sim \, 0.1
\label{scalesst}
\ee
where now all the volumes are measured in string units and in the string frame. In terms of the Einstein frame we have that
\be
\hat L_{X_6} \, \sim\, 10^{15/24} \quad \quad \hat L_{\Pi_3} \, \sim\, 10^{11/24} \quad \quad g_s^2\, \sim \, 0.1
\label{scalesE}
\ee
and so plugging these values in the above expressions we find that the inflaton mass in Planck units is given by
\be
\kappa_4\, m_{\rm inf}\, \sim \, Z^{-p/2}_{\text{D6}}  10^{-35/10}
\ee
Hence one recovers the standard value of $m_{\rm inf} \sim 10^{13} GeV$ by considering a warp factor of the order $Z_{\rm D6}^p \sim 10^3$. Finally, plugging the values (\ref{scalesE}) into (\ref{ratioTinf}) we find
\be
m_{T^\a}^2\, \sim\, 10^{-1} n^2 Z^p_{\text{D6}}\, m_{\rm inf}^2 \quad \quad \Raw \quad \quad m_{T^\a}\, \sim\, 10 n \, m_{\rm inf}
\label{ratioTinf2}
\ee
where we have plugged the above value for the warp factor. Therefore by setting the flux quanta of  the order $n \sim 10$ or higher we find an acceptable hierarchy between the masses induced by the flux superpotential and that of the inflaton candidate.

\section{The B-field potential for large field values}
\label{s:largeb}

Out of the two type IIA scenarios discussed in the last section, one of them has a technical advantage over the other. Indeed, for the scenario in section \ref{Bfield}, in which the inflaton is a B-field axion, we are able to compute $\a'$ corrections to the supergravity effective potential by using the DBI potential obtained in section \ref{s:lifting}. More precisely, if we take the supergravity effective potential (\ref{Vfinal}) and we evaluate it at $\Phi=0$ then we have that it reduces to
\be
V\, =\, \frac{\pi}{\kappa_4^4} e^K K^{\Phi\bar\Phi}|T|^2\, =\, \frac{1}{\kappa_4^4} \frac{\pi g_s^{-1/2}}{8(\hat{V}_{X_6}^{E,0})^3}  \frac{ K^{\Phi\bar \Phi} \left( b^2 + j^2 \right)}{1 - 2 K_{T\bar T}^0 j^2}
\label{sugrapotT}
\ee
where $\hat{V}_{X_6}^{E, 0}$ is the compactification volume in the Einstein frame and $K_{T\bar{T}}^0$ the the kinetic terms for the complex field $T$ evaluated at $j=0$. As in (\ref{Tsquared}) $b$ stands for the unit period axion and $j$ for its saxion partner. At large values of $|T|$ this potential is replaced by one obtained from the DBI action, namely the square-root potential of eq.(\ref{finalpot}). In general, evaluating of such potential will depend on the specific geometry of the three-cycle $\Pi_3$ wrapped by the D6-brane. Let us however take the simplifying assumption that the quantity $\rho^2/\om^2$ inside the square bracket is constant over $\Pi_3$ and independent of $j$. In that case the potential can be approximated by  
\be
V_{\rm D6} \, \simeq \,   \frac{1}{\kappa_4^4} \frac{g_s^{3/4}\hat{V}_{\Pi_3}^0}{2\pi (\hat{V}_{X_6}^0)^2} \frac{1}{1 - 2 K_{T\bar T}^0 j^2}  \left( \sqrt{ 1 + \frac{\pi^2 K^{\Phi\bar \Phi}}{2 g_s^{5/4} \hat{V}_{X_6}^0 \hat{V}_{\Pi_3}^0}\, (b^2 + j^2) } - 1 \right)
\label{DBIpotT}
\ee
which clearly reduces to (\ref{sugrapotT}) for small values of $|T|$. Notice that in this limit the kinetic terms for $b$ and $j$ are not canonical but given by 
\be
K_{T\bar{T}}\, =\, K_{T\bar{T}}^0 \cdot \frac{1 + 2 K_{T\bar{T}}^0 j^2}{(1- 2 K_{T\bar{T}}^0 j^2)^2}
\ee
Since this kinetic term arises from a bulk integral computed at an arbitrary point of the K\"ahler moduli space, we will assume that it receives no corrections when we climb up along the inflationary potential.\footnote{It would be interesting to explore if some corrections could be induced through backreaction effects.} Therefore the only effects of $\alpha'$ corrections to the inflationary dynamics appears through the square-root behaviour of the potential (\ref{DBIpotT}).  

Notice that the corrected potential (\ref{DBIpotT}) only includes the dependence of one of the two complex fields $(T, \Phi)$ of the inflationary sector. Ideally one would like to have a corrected potential for both of the complex fields in order to analyse the stability of the inflationary trajectory (\ref{infTB}). Nevertheless, by the analysis of the previous section and Appendix \ref{ap:4dsugra} we have seen that the inflaton $b$ and its saxionic partner $j$ are the two lightest fields of the system in the supergravity limit. If we assume that such hierarchy of scales is still valid at large field values we may set $\Phi=0$ and then recover the potential (\ref{DBIpotT}). In the following we will take such approach and analyse the dynamics for the fields $b$ and $j$ from (\ref{DBIpotT}). In fact, in section \ref{s:saxion} we will see that this $\alpha'$-corrected potential exactly reproduces the saxion mass estimate obtained in (\ref{massesB}). Therefore along the inflationary trajectory it makes sense to set $j=0$ and study the single field inflationary potential for $b$, as we will do in the following.

\subsection{Slow roll parameters for large inflaton vevs}

Along the inflationary trajectory (\ref{infTB}) the $\alpha'$ corrected inflationary potential for the B-field $b$ can be taken directly from (\ref{DBIpotB}) by taking $\Pi_3$ to be an special Lagrangian. By making the simplifying assumption that $\rho^2$ is constant along the three-cycle $\Pi_3$ (or equivalently that $\rho \wedge *_3 \rho$ is harmonic on $\Pi_3$) we recover a potential of the form\footnote{Interestingly, such potential form is also recovered in one of the single field limit cases of \cite{Ibanez:2014swa} after the fields have been canonically normalised. See \cite{Bielleman:2015lka} for more details.}
\be
V\, \simeq \, \g \left( \sqrt{ 1 + \d \left(\frac{\phi_b}{M_{\rm pl}}\right)^2} -1 \right) M_{\rm pl}^4
\label{sqrtpotb}
\ee
where $\phi_b = M_{\rm pl} \sqrt{2K_{T\bar{T}}^0} \, b$ is the canonically normalised B-field in the scenario of section \ref{Bfield}. Alternatively one may take the limit $j\raw 0$ in (\ref{DBIpotT}). In both cases one obtains that the dimensionless parameters $\b$ and $\g$ are given by
\bea
\label{cparam}
\g & \sim &  \frac{1}{2\pi}  g_s^{3/4}  \frac{\hat{V}_{\Pi_3}^{E, 0}}{(\hat{V}_{X_6}^{E, 0})^2}\, \sim\, 10^{-7}\\
\d^{-1} & \sim & \frac{4}{\pi^2} g_s^{5/4} K_{\Phi\bar\Phi} K_{T\bar T}^0 \hat{V}_{\Pi_3}^{E, 0} \hat{V}_{X_6}^{E, 0} \, \sim\, 10^{2}
\label{aparam}
\eea
where we have estimated the value of these parameters by plugging the values (\ref{scalesE}) as well as $Z_{\rm D6}^p \sim 10^3$ used in the previous section. As these values may slightly vary from one model to another, in particular $\beta$ due to the approximations that we have taken, let us take a phenomenological approach and analyse the potential (\ref{sqrtpotb}) for the parameter range
\be
\d \, \sim \, 10^{-1} - 10^{-3} \quad \quad \quad \sqrt{\d\g} \, \sim \, 10^{-4} -10^{-5}
\label{range}
\ee

Given this single field inflationary potential one may compute the cosmological parameters associated to the range (\ref{range}). In particular one finds that  slow-roll inflation typically occurs for $1.4 M_{\rm pl} < \phi_b  < 13-15 M_{\rm pl}$ for 60 efolds, and for $1.4 M_{\rm pl} < \phi_b < 12-14 M_{\rm pl}$ for 50 efolds, the precise upper limit $\phi_{b\,*}$ depending on the value of $\d$. Since the $b \sim K_{T\bar{T}}^{1/2} \phi_b$ we find that the number of periods that the axion must undertake is of order $10^2$. 

In general, cosmological parameters of the model are mostly sensitive to the value of $\d$, which interpolates between a model of quadratic chaotic inflation ($\d \sim 10^{-3}$) and  linear chaotic inflation ($\d \sim 10^{-1}$). 
%
\begin{figure}[h!]
\begin{center}
{\includegraphics[width=0.49\textwidth]{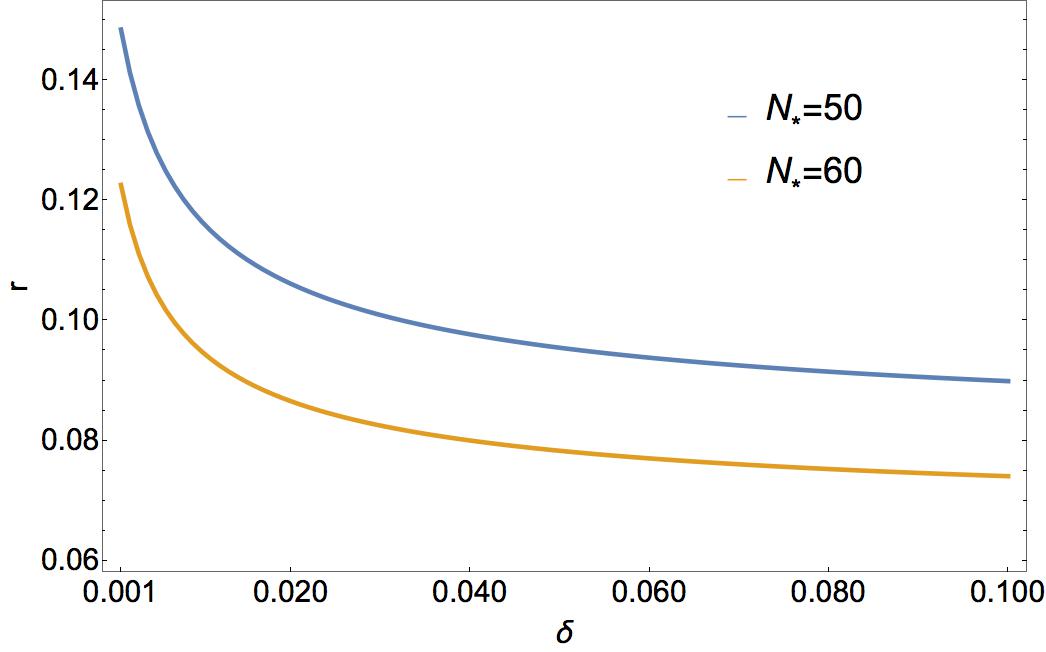}} 
{\includegraphics[width=0.49\textwidth]{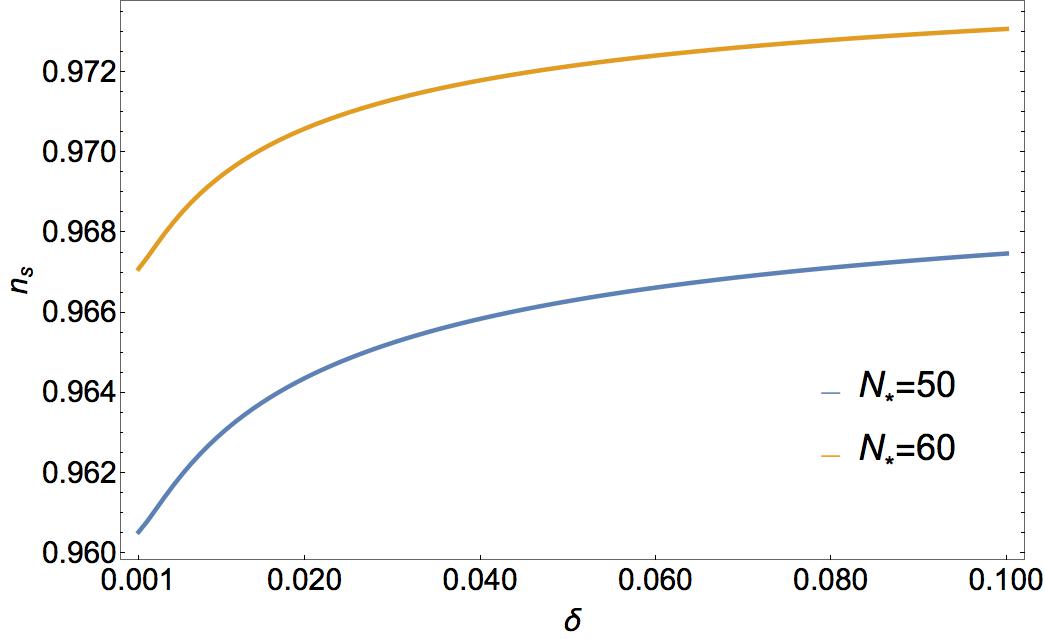}} 
\end{center}
\vspace{-15pt}
\caption{Tensor-to-scalar ratio (left) and spectral index (right) in terms of $\d$.}
\label{rns}
\end{figure}

\begin{figure}[!h]
\begin{center}
\includegraphics[scale=0.44]{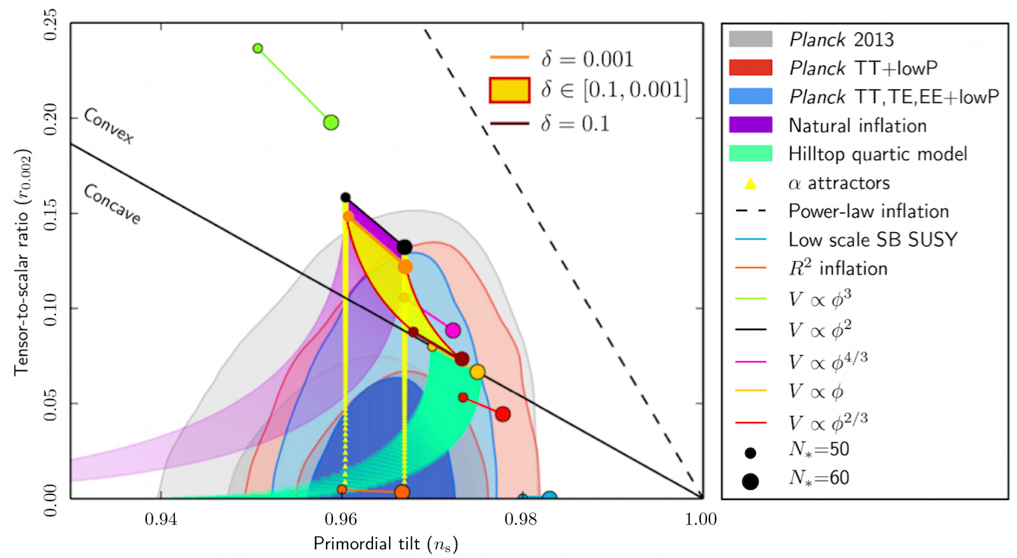}	
\end{center}	
\protect\caption{Primordial tilt $n_{s}$ vs tensor-to-scalar ratio $r$ superimposed by the plot given by Planck Collaboration (2015) \cite{Ade:2015tva}. The yellow area shows the region of parameters covered by the potential (\ref{sqrtpotb}) for the parameter range $\d \sim 10^{-1} - 10^{-3}$.}
\label{nsrplanck}
\end{figure}

In figure \ref{rns} we display the tensor-to-scalar ratio and the spectral index in terms of the parameter $\d$, for the number of efolds $N_* = 50$ (blue line) and $N_* =60$ (red line). Their behaviour can be understood in terms of an interpolation from quadratic to linear inflation as we increase the value of $\d$. Such interpolation is also illustrated by plotting one cosmological parameter in terms of the other and superimposing the result on the plot recently given by the Planck collaboration \cite{Ade:2015tva}, as we do in figure \ref{nsrplanck}.\footnote{This interpolation is also recovered in the context of field theory in \cite{Kannike:2015kda}, up to UV completion effects.}

\subsection{Stability bounds on the DBI potential}
\label{s:saxion}

Given the $\alpha'$-corrected potential (\ref{DBIpotT}) we may revisit the computation that, in the supergravity approximation, led us to the estimate (\ref{massesB}) for the mass of the saxion $j$ along the inflationary trajectory. For this it is useful to rewrite the potential (\ref{DBIpotT}) as
\be
V\, \simeq \,  \frac{\g}{1 - 2 K_{T\bar T}^0 j^2} \left( \sqrt{ 1 + \lam \left( b^2 + j^2 \right)} -1 \right) M_{\rm pl}^4
\label{sqrtpotbj}
\ee
where $\g$ is defined as in (\ref{cparam}) and
\be
\lam \, =\,  2K_{T\bar{T}}^0\, \d
\label{dparam}
\ee
Let us now repeat the computation below eq.(\ref{ap:saxionmass}) for the current potential. As in there we have that
\be
m_{\rm saxion}^{2}\mid_{\rm Traj}\, =\, \frac{1}{2K_{T\bar T}} \p_{j}^{2} V\mid_{\rm Traj} 
\label{saxionmass}
\ee
with the trajectory given by (\ref{infTB}) which implies $j=0$. We then find
\bea
m_{\rm saxion}^{2}\mid_{\rm Traj} & = & \g \left(\frac{\lam/(2K_{T\bar{T}}^0)}{ \sqrt{ 1 + \lam \, b^2 }} +  2\left[ \sqrt{ 1 + \lam \, b^2 } -1 \right]\right) \\
 & = & \g \left[ \sqrt{ 1 + \lam \, b^2 } -1 \right]  \left(\frac{2}{1+\frac{1}{\sqrt{ 1 + \lam \, b^2 } }} \, \epsilon + 2 \right) \\
 & = & 3H^2 \left(\frac{2}{1+\frac{1}{\sqrt{ 1 + \lam \, b^2 } }}\, \epsilon + 2\right)
\eea
where 
\be
\eps\, =\, \frac{1}{4K_{T\bar{T}}^{0}} \left(\frac{b\, \lam  }{\sqrt{1+\lam b^2}(\sqrt{1+\lam b^2} -1)} \right)^2
\ee
During inflation $\eps \ll 1$ and so we can neglect the piece proportional to it. Then we obtain
\be
m_{\rm saxion}^{2}\mid_{\rm Traj}\, \simeq\, 6 H^2
\ee
in agreement with the supergravity result (\ref{saxionm1fin}).

\section{Conclusions and Outlook}
\label{s:conclu}

In this paper we developed the proposal made in \cite{Escobar:2015fda} to realise models of large field inflation by including D-branes that generate a bilinear superpotential of the form (\ref{introWinf}). For concreteness we have focused on type IIA compactifications with D6-branes, although most of our results are also valid in dual setups like type IIB/F-theory compactifications with 7-branes. Since the superpotential (\ref{introWinf}) has been used extensively in the 4d supergravity literature to build models of large field inflation, we have considered compactifications whose inflaton sector resembles such supergravity models as much as possible. This is a non-trivial task, since type IIA compactifications contain many extra scalars beyond the inflaton sector that need to be stabilised well above the Hubble scale, and that mix with the fields in the superpotential (\ref{introWinf}) via a specific K\"ahler potential. Nevertheless, by imposing some assumptions on the compactification background we have managed to recover an inflaton sector quite similar to those in \cite{Kawasaki:2000yn,Kallosh:2010ug,Kallosh:2010xz,Kallosh:2014vja,Nakayama:2014wpa} after all the heavier scalars have been stabilised. The 4d supergravity description is however only valid for small inflaton vevs. For trans-Planckian vevs, $\alpha'$ effects may induce important corrections to the scalar potential. We have been able to compute such corrections for the scenario where the inflaton descends from a B-field component, obtaining a flattened potential with a linear behaviour for large inflaton values. Such flattening of the potential has a non-trivial effect on the cosmological parameters of the model. In particular it lowers the value of the tensor-to-scalar ratio with respect to the quadratic potential of the related 4d supergravity models, allowing to easily fit the resulting ratio  within current experimental bounds. 

Based on our results, there is a number of open problems and further developments that need to be addressed in order to construct concrete models and obtain precise predictions out of them. For instance, one important development would be to construct explicit examples of special Lagrangian three-cycles that contain two-cycles which are non-trivial in the bulk geometry. As explained in \cite{Marchesano:2014iea} and in the main text, such topological condition is necessary to generate the superpotential (\ref{introWinf}) and therefore the scalar potential for the inflation system. While examples of these three-cycles can be obtained in simple toroidal orbifold geometries \cite{Font:2006na}, it would be desirable to gain a better understanding of their properties by constructing them in smooth Calabi-Yau geometries, perhaps by using the techniques in \cite{Uranga:2002pg,Palti:2009bt,Apruzzi:2014dza}. In particular, it would be very interesting to compute the DBI potential for such explicit examples. One could then see if the assumptions made to arrive to the square-root potential (\ref{DBIpotT}) are realised in practice or if on the contrary a different sort of of flattened potential is obtained. 

We have also made a number of assumptions when embedding our inflaton system in a type IIA moduli stabilisation scenario, which would be interesting to realise in explicit compactification geometries. Most of these assumptions should be easy to implement in general, as the toy model in appendix \ref{WLtoy} illustrates, but others may be less generic in the current landscape of type IIA flux compactifications. One particular condition that we need to impose is that the chiral field containing the inflaton does not appear in the superpotential used for moduli stabilisation. This is easy to implement for the Wilson line scenario of section \ref{WLine}, but it less obvious for the B-field scenario of section \ref{Bfield}, at least for the scenarios of type IIA moduli stabilisation that have so far been explored in the literature. Nevertheless, one would expect that this condition is achievable upon applying mirror symmetry to the type IIB flux compactifications discussed in \cite{Blumenhagen:2014nba}, since they contain examples where the superpotential does not depend on a particular complex structure modulus. Alternatively, one may directly construct this scenario in the context of type IIB flux compactifications, where a superpotential of the form (\ref{introWinf}) is generated by considering D7-branes with three-cycles that are also homologically non-trivial in the bulk \cite{Marchesano:2014iea}. Finally, one may consider dropping some of the assumptions made in section \ref{s:scenarios} and see if the effective potential for the inflaton system is still suitable to drive inflation. 

A more ambitious but equally important future direction would be to compute the $\a'$ corrected potential for the whole inflaton system, and not only for the closed string fields as we have done in section \ref{s:largeb}. Having such corrected potential would allow to check if the stability conditions for the inflationary trajectory obtained in the supergravity regime are still valid for large inflaton vevs. Moreover, it would allow to check if the inflationary potential for the Wilson line is also flattened at large field values, as has been observed for the B-field scenario. 

We hope to come back to these and other open problems in the near future. In any event, we believe that our results throw some light into the possibilities of string compactifications to achieve successful models of large field inflation.

\bigskip

\bigskip

\centerline{\bf \large Acknowledgments}

\bigskip

We would like to thank Thomas Grimm, Luis Ib\'a\~nez, Eran Palti, Francisco Pedro, Hagen Triendl, Timo Weigand and specially Clemens Wieck for useful discussions. 
This work has been partially supported by the grant FPA2012-32828 from the MINECO, the REA grant agreement PCIG10-GA-2011-304023 from the People Programme of FP7 (Marie Curie Action), the ERC Advanced Grant SPLE under contract ERC-2012-ADG-20120216-320421 and the grant SEV-2012-0249 of the ``Centro de Excelencia Severo Ochoa" Programme. In additon D. E. and A.L. are respectively supported through the FPI grants SVP-2014-068283 and SVP-2013-067949, F.M. is supported by the Ram\'on y Cajal programme through the grant RYC-2009-05096 and D.R is supported by a grant from the Max Planck Society. F.M. would like to thank HKUST IAS, the CERN TH Division, the Aspen Center for Theoretical Physics (supported through the NSF grant PHY-1066293) and UW-Madison for hospitality and support during the completion of this work.

\clearpage

\appendix


\section{D6-brane DBI computation}
\label{ap:DBI}

In the setup of section \ref{s:lifting}, the potential felt by the B-field axion is due to the energy increase on a D6-brane worldvolume when we move in K\"ahler moduli space. As discussed in that section and in \cite{Marchesano:2014iea}, near the point where the D6-brane is BPS we can understand such energy increase in terms of an F-term potential generated by the superpotential (\ref{Winf}). However, inflation will occur far away from the point where the 4d supergravity description is valid, and therefore we need to compute the potential for the K\"ahler modulus directly from the DBI action. In the following we will perform such computation, generalising the results of \cite{kt11} to the case where the complexified K\"ahler form $J_c$ has a non-trivial pull-back on the three-cycle $\Pi_3$ wrapped by the D6-brane. In the process we will derive several identities that will be used in section \ref{s:lifting}.

The DBI action for a D6-brane is given by
\be
S_{DBI}=-\mu_6 g_s^{-1}{\rm STr}\left (\int d^7x\sqrt{-{\rm det}\left (P[E]_{MN}-\sig F_{MN}\right )} \right )
\label{DBID6ap}
\ee
with
\be
E=g_s^{1/2}g+B \qquad  \qquad \mu_6=2\pi/l_s^7 \qquad  \qquad \sig=\frac{l_s^2}{2\pi}
\ee
and where $g$ is the ten-dimensional metric in the Einstein frame.

This action depends explicitly on the gauge field living on the worldvolume of the D6-brane and implicitly on the scalars that parametrise its position in the bulk. To make the dependence on the latter manifest, consider a D6-brane on $\IR^{1,3}\times \Pi_3$ where $\Pi_3$ is a three-cycle of the compactification space $X_6$. We first choose a reference submanifold $\hat\Pi$ and introduce coordinates on the bulk $x^\Sigma=(x^M,x^m)$ such that the embedding of $\hat \Pi$ is given by $x^m=0$. A normal vector to $\hat \Pi$ is then given by $N^\Sigma=(0,\sig f^m(x^M))$ for any function $f^m$. Now we may parametrise the embedding of $\Pi_3$ as a deformation of $\hat \Pi$ along a normal vector $N^\Sigma$, namely, the embedding map for $\Pi_3$ is given by $g^\Sigma=(x^M,\sig f^m(x^M))$. We see that the general expression for the pullback
\be
P[E]_{MN}\, =\, \frac{\p g^\Sigma}{\p x^M}\frac{\p g^\Lambda}{\p x^N}E_{\Sigma\Lambda}\,,
\ee
reduces in this case to
\be
P[E]_{MN}\, =\, E_{MN} + \sig \p_Mf^m E_{mN} + \sig \p_N f^n E_{Mn} + \sig^2 \p_M f^m\p_N f^n E_{mn},
\label{PBE}
\ee
which can be plugged into \eqref{DBID6ap}. One then finds that
\be
\det (P[E]_{MN}-\sig F_{MN})=g_s^{7/2}\det \left (\begin{array}{cc}
A &  B\\
C & D  \end{array}\right )
\ee
with 
\bea\nonumber
A=\eta_{\mu\nu} - \sig g_s^{-1/2}  F_{\mu\nu} +\sig^2 \p_\mu f^m \p_\nu f^ne_{mn}\, &&\quad B= \sig g_s^{-1/2} \p_\mu \Phi_b^- + \sig^2 \p_\mu f^m\p_bf^ne_{mn}\\
C= \sig g_s^{-1/2} \p_\nu \Phi_a^+ + \sig^2 \p_af^m\p_\nu f^ne_{mn}\, && \quad D= P[g]_{ab}+g_s^{-1/2}\CF_{ab}\,,
\eea
where we defined $\CF=P[B]-\sig F$ and
\be
e=g+g_s^{-1/2}B,\qquad \quad \Phi_a^\pm\, =\, \pm A_a + (g_s^{1/2} g_{an} \pm B_{an}) f^n
\ee
and $\mu, \nu$ run over the 4d coordinates and $a,b$ over the three coordinates of $\Pi_3$. 
Using the following identities
\bea\label{matiden}
\begin{split}
\det \left (\begin{array}{cc}
A&B\\
C&D\end{array}\right )&=\det A\det (D-CA^{-1}B)\\
\det(\mathbb I+\eps A)&=1+\eps\,\tr A+\frac{\eps^2}{2}((\tr A)^2-\tr A^2)+\mathcal O(\eps^3)\\
\det(\mathbb I_3+ A_{3\times 3})&=1+\tr A+\frac{1}{2}((\tr A)^2-\tr A^2)+ \det A
\end{split}
\eea
we can expand the action to second order in derivatives along the non-compact directions. This yields
\bea
\begin{split}\label{apDBIexp}
S_{DBI}&=-\int d^7x\, \CV \left [ 1+\frac{\sig^2}{2}\left ( \frac{1}{2g_s}F_{\mu\nu}F^{\mu\nu}+ \p_\mu f^m\p^\mu f^n g_{mn} \right .\right .\\
&\left .\left .\qquad - (D^{-1})^{ab}(g_s^{-1/2}\p_\mu\Phi_b^++\sig\p_bf^m\p_\mu f^ne_{mn})(g_s^{-1/2}\p^\mu\Phi_a^-+\sig\p_af^p\p^\mu f^qe_{qp})\right )\right ]
\end{split}
\eea
where we defined
\be\label{defM}
\CV =\mu_6g_s^{3/4} \sqrt{\det D}.
\ee
We have now managed to make the dependence on the embedding $f^m$ manifest. The usual procedure at this point is to compute the equations of motion, split the 7d fields in internal and external components and use separation of variables which allows to obtain different equations for the internal and external fields. The equations for the internal wavefunction correspond to an eigenvalue problem for an elliptic operator (a generalisation of the Laplace operator) and the eigenfunctions yield the tower of KK replicas. This is a very complicated problem for the action \eqref{apDBIexp} but, fortunately, it is more than we need since we are only interested in the lowest KK mode. Thus, we may write 
\be\label{decom}
A_a=A_a^0+ \xi(x^\mu)\zeta_a(x^b)\,,\qquad \qquad  f^m=f^m_0(x^a)+\varphi(x^\mu)X^m(x^a)
\ee
where $\zeta_a$ and $X^m$ are the internal wavefunctions and $A_a^0,\,f^m_0$ are background profiles for the gauge field and embedding of the brane. In the case where the D6-brane is in a supersymmetric configuration, $\zeta_a$ is a harmonic one-form and we have that $g_s^{1/2}X^m=J^{am}\zeta_a$. Plugging the ansatz \eqref{decom} in the action and expanding to quadratic order in the open string fields $\xi,\,\varphi$ we find that\footnote{Here we performed the Weyl rescaling \eqref{weylr} to bring the action into the 4d Einstein frame.}
\bea
\begin{split}\label{DBIexpap}
S_{DBI}&=-\int d^7x\, \CV_0\left [ \frac{4\tilde Q}{\hat V_{X_6}^2} +\frac{\sig^2}{4g_s}F_{\mu\nu}F^{\mu\nu}+ \frac{1}{2}(\p_\mu \varphi\,\,\,\p_\mu\xi )\,\mathbf{\tilde M}\left (\begin{array}{c}\p^\mu\varphi\\\p^\mu\xi\end{array}\right ) \right ]
\end{split}
\eea
with $\CV_0=\mu_6g_s^{3/4} \sqrt{\det D_0}$, $D_0$ being $D$ with $\varphi$ and $\xi$ equal to zero. We have
\be
\tilde Q=1+\frac{\sig}{2} Q_1+\frac{\sig^2}{8}Q_1^2+\frac{\sig^2}{4} Q_2
\ee
with
\bea\label{cristo}
Q_1&=& \varphi M^{(ab)}(\mathcal L_X\tilde g)_{ab} +  g_s^{-1/2} M^{[ab]}(\xi d\zeta-\varphi \mathcal L_X \tilde B)_{ab}\\
Q_2&=& \varphi^2 M^{(ab)} (\mathcal L_X \mathcal L_X \tilde g)_{ab}-g_s^{1/2} \varphi^2 M^{[ab]}(\mathcal L_X \mathcal L_X \tilde B)_{ab} -\\\nonumber
&& -  M^{ab}M^{cd}( \varphi \mathcal L_X\tilde g+g_s^{-1/2}(\varphi \mathcal L_X\tilde B - \xi d\zeta))_{bc}( \varphi \mathcal L_X\tilde g+g_s^{-1/2}(\varphi \mathcal L_X\tilde B - \xi d\zeta))_{da}
\eea
where we defined\footnote{The matrix $M$ appears oftentimes in the literature of bosonic and fermionic actions, see e.g. \cite{Martucci:2005rb,Marchesano:2010bs}.}
\be
M^{ab}=(D_0^{-1})^{ab}.
\ee
The Lie derivative is given by $(\mathcal L_X T)_{ab}=X^m\p_mT_{ab}+ \p_a X^m T_{mb}+ \p_bX^mT_{am}$ with
\bea\nonumber
\tilde g_{mb}=g_{mb}+\sig\p_bf_0^ng_{mn}&\qquad&  \tilde g_{ab}=g_{ab}+\sig\p_a f_0^mg_{mb}+\sig\p_b f_0^mg_{am}+\sig^2\p_af_0^m\p_bf_0^ng_{mn}\\\nonumber
 \tilde B_{mb}=B_{mb}+\sig\p_bf_0^nB_{mn}&\qquad& \tilde B_{ab}=B_{ab}+\sig\p_a f_0^mB_{mb}+\sig\p_b f_0^mB_{am}+\sig^2\p_af_0^m\p_bf_0^nB_{mn}\,.
\eea
Finally, the kinetic terms read,
\bea\label{kinkin}
\begin{split}
\mathbf{\tilde M}_{\varphi\varphi}&=\frac{2\sig^2}{\hat{V}_{X_6}}X^mX^n \left (g_{mn}-M^{(ab)}(\tilde g_{ma}+g_s^{-1/2}\tilde B_{ma})(\tilde g_{nb}-g_s^{-1/2}\tilde B_{nb})\right )\\
\mathbf{\tilde M}_{\varphi\xi}&=- \frac{2\sig^2}{g^{1/2}_s\hat{V}_{X_6}}X^m\zeta_b(M^{[ab]}\tilde g_{ma}+M^{(ab)}g_s^{-1/2}\tilde B_{ma})  \\
\mathbf{\tilde M}_{\xi\xi}&=\frac{2\sig^2}{g_s\hat{V}_{X_6}}M^{(ab)}\zeta_a\zeta_b\,.
\end{split}
\eea
The matrix $M^{ab}$ above can be computed explicitly using the fact that it is a $3\times 3$ matrix which yields
\be\label{Mexp}
M^{(ab)}=\hat g^{ab}+\frac{\CF^{ac}\CF_{c}{}^b}{g_sW_F},\qquad 
M^{[ab]}=-\frac{\CF^{ab}}{g_s^{1/2}W_F},\qquad W_F={1+\frac{1}{2g_s}\mathcal F_{ab}\mathcal F^{ab}}.
\ee
where $\hat g^{ab}$ is the inverse of $P[g]_{ab}$ and we have used $\hat g^{ab}$ to raise the indices of $\mathcal F$ in the expression above.

\subsubsection*{Kinetic terms}

Once we integrate over the internal space, the kinetic term for the vector in the action \eqref{DBIexpap} is
\be
S_{4d}^{v}=-\frac{\mu_6 \sig^2}{2}\int d^3xg_s^{-1/4}\sqrt{\det (P[g]+g_s^{-1/2}\CF)}\int F\wedge *F\,.
\ee
For a supersymmetric configuration with constant dilaton we have $\CF=0$ and the above reduces to
\be
S_{4d}^{v}=-\frac{\mu_6 \sig^2 V_{\Pi_0}}{2g_s^{1/4}}\int F\wedge *F\,,
\ee
where $V_{\Pi_0}$ is the volume of $\Pi_0$, the cycle defined by the embedding $f^m_0$.

The kinetic term for the scalars is
\be\label{sca}
S_{4d}^{s}=-\frac{1 }{2}\int d^4x\,(\p_\mu \varphi\,\,\,\p_\mu\xi )\,\mathbf{M}\left (\begin{array}{c}\p^\mu\varphi\\\p^\mu\xi\end{array}\right ).
\ee
with
\be
\mathbf{M}=\mu_6\int d^3x\, g_s^{3/4}\sqrt{\det (P[g]+g_s^{-1/2}\CF)}\, \tilde{\mathbf M}.
\ee
For the case of a D6-brane wrapping a special Lagrangian cycle with non-vanishing worldvolume flux $\CF$, which is the inflationary trajectory that in section \ref{s:largeb} we have that the kinetic terms for the open string fields are
\be\label{kinkin2}
\mathbf{\tilde M}_{\varphi\varphi}= \frac{2\sig^2}{g_s\hat{V}_{X_6}}\zeta_a\zeta_b\,(g^{ab}+M^{(ab)}\eta^2),  \quad \mathbf{\tilde M}_{\varphi\xi}= - \frac{2\sig^2}{g_s\hat{V}_{X_6}}\eta\, M^{(ab)} \zeta_a\zeta_b,  \quad \mathbf{\tilde M}_{\xi\xi}=\frac{2\sig^2}{g_s\hat{V}_{X_6}}M^{(ab)}\zeta_a\zeta_b\,,
\ee
where we defined $\eta\,\zeta_a=X^m\tilde B_{ma}$ with $\eta\in\IR$ and used that $g_s^{1/2}X^m=J^{am}\zeta_a$. Finally, setting $\CF=0$ yields a supersymmetric configuration and the above simplifies to
\be\label{kinkin3}
\mathbf{\tilde M}_{\varphi\varphi}= \frac{2\sig^2}{g_s\hat{V}_{X_6}}\zeta\wedge *\zeta\,(1+\eta^2),  \qquad \mathbf{\tilde M}_{\varphi\xi}= - \frac{2\sig^2}{g_s\hat{V}_{X_6}}\eta\,  \zeta\wedge *\zeta,  \qquad \mathbf{\tilde M}_{\xi\xi}=\frac{2\sig^2}{g_s\hat{V}_{X_6}}\zeta\wedge *\zeta\,.
\ee
so we can write eq.\eqref{sca} as
\be\label{scasusy}
S_{4d}^{s}=- \frac{1}{2\kappa_4^2} \frac{1}{8g_s^{1/4}\hat V_{X_6}} \frac{1}{l_s^3}\int  \zeta\wedge *\zeta \int d\Phi\wedge*d\bar\Phi\,,
\ee
with $\Phi=\frac{l_s}{\pi}(\xi - \eta\varphi-i\varphi)$, which matches the result obtained in \cite{kt11} and, using the dictionary of footnote \ref{foot}, corresponds to eq.(\ref{kinxixi}) in the main text.

\subsubsection*{Potential}

Let us now turn discuss the potential term in \eqref{DBIexpap}. It reads
\be\label{DBIpot}
S^{p}_{DBI}=-\frac{4\mu_6}{\hat V_{X_6}^2}\int d^7x\, g_s^{3/4}\, \tilde Q  \sqrt{\det (P[g]_{ab}+g_s^{-1/2}\CF_{ab})}\,
\ee
where the pullback is over the cycle $\Pi_3$ with embedding function $g^\Sigma=(x^M,\sig f_0^m(x^M))$ and is therefore independent of $\varphi,\,\xi$. We analyse first the potential for the closed string moduli and subsequently the one for the open string fields.

\paragraph{Closed string fields.}

The term that is independent of the open string fields can be written as
\be\label{DBIpotp}
S^{p,c}_{DBI}=-\frac{4\mu_6}{\hat V_{X_6}^2}\int d^4x\,d{\rm vol}_{\Pi}\, g_s^{3/4} \sqrt{1+\frac{1}{2g_s}\CF_{ab}\CF^{ab}} 
\ee
where we have used \eqref{matiden} and the induced metric $P[g]$ to raise the indices of $\CF$. We clearly see that the potential energy grows whenever there is a non-vanishing field strength $\mathcal F$ on the D6-brane, which fits nicely with (the real part of) the supersymmetry condition \eqref{Fterm}. However, it is not so clear how to see the special Lagrangian condition from the potential above. In order to do so, we need to rewrite the volume form in $\Pi_3$ in a convenient way, as follows.

Consider a point $p\in\Pi_3$ and choose coordinates in the bulk such that the metric, K\"ahler form and holomorphic 3-form are canonical at $p$,
\bea\begin{split}
\Omega & \,= \,dz_1 \wedge dz_2 \wedge dz_3 \quad \quad dz_i = dx_i + i dx_{i+3}\\
J & \,= \, dx_1 \wedge dx_4 +  dx_2 \wedge dx_5 +  dx_3 \wedge dx_6.
\end{split}
\eea
Then, the pullbacks of $J$ and $\Omega$ on $\Pi_3$ are given by
\bea\begin{split}
P[g]_{ab} \, =\, \d_{ab} + \p_a f_c \p_b f_c\,, &\quad \quad & g \, =\, \Id_3  + AA^t\\\label{POmega}
P[\Omega]_{abc}\, =\, \text{det} \left( \Id_3 + i A \right) \\
P[J]_{ab}\, =\, \p_af_b - \p_bf_a\,, &\quad \quad & J\, =\, A - A^t
\end{split}
\eea
where we have defined 
\be
A_{ab} \, =\, \p_a f_b\,,\qquad \qquad f_a=J_{ma}f_0^m\,.
\ee
Using eq.\eqref{POmega} we have that
\be
 (\re\, P[\Omega])^2  + (\im\, P[\Omega])^2 \, =\, (\text{det}\, g)^{-1}  \text{det} \left( \Id_3 + i A \right) \left( \Id_3 - i A^t \right), 
\ee
where we defined $Q_p^2=\frac{1}{p!}Q_{a_1\cdots a_p}Q^{a_1\cdots a_p}$ for any $p$-form $Q_p$.
Then one has that
\be\nonumber
\text{det} \left( \Id_3 + i A \right) \left( \Id_3 - i A^t \right)  = \text{det} ( g + i J) =  \text{det} g\,  \text{det} ( \Id_3 + i g^{-1}J) = \text{det} g  \left( 1 + \oh \tr  (g^{-1}J)^2 \right)
\ee
where in the last step we used \eqref{matiden} and the fact that $J$ is antisymmetric.
Putting both results together we have that 
\be
 P[J]^2 +  (\re\, P[ \Omega])^2 +  (\im\, P[\Omega])^2 \, =\, 1\,.
\ee
Notice that we arrived at this equation in a particular set of coordinates but, since it is tensorial, it is true in any coordinate system. Finally, using that 
\be
d\text{vol}_{\Pi}\, =\, \frac{ \re\,P[ \Omega]}{\sqrt{(\re\, P[\Omega])^2}}\,,
\ee
we find that we can rewrite \eqref{DBIpotp} as
\be\label{DBIpot2}
S^{p,c}_{DBI}=-4\mu_6\int dx^4 \int_{\Pi_3} \re\,\Omega \,\frac{g_s^{3/4}}{\hat V_{X_6}^2} \sqrt{1+\frac{P[J]^2+(\im\,P[\Omega])^2+g_s^{-1}\mathcal F^2}{(\re\,P[\Omega])^2}}\,.
\ee
 If the brane is close to being supersymmetric ($(\re\, P[\Omega])^2\simeq 1$) we can Taylor expand the square root and find that
\be\label{DBIpot3}
S^{p,c}_{DBI}\simeq-4\mu_6\int dx^4 \int_{\Pi_3} \re\,\Omega \,\frac{g_s^{3/4}}{\hat V_{X_6}^2} \left (1+\oh P[J]^2+\oh (\im\,P[\Omega])^2+\frac{1}{2g_s} \mathcal F^2\right )\,,
\ee
which shows that to preserve supersymmetry we need to impose \eqref{Fterm} and \eqref{Dterm}.

\paragraph{Open string fields.} The terms of the potential that depend on the open string moduli $\varphi$ and $\,\xi$ are contained in the polynomial $\tilde Q$, which contains one linear term (\ref{cristo}) in the fields and another one which is quadratic. Whenever the linear term is non-vanishing the open string fields do not sit at the minimum of the potential and do not satisfy their equations of motion. In order to have a better intuition on how this affects the D-brane dynamics let us consider the case where $\Pi_3$ is a special Lagrangian and so we have that 
\be
\iota_X \tilde B\, =\, \eta\, \zeta
\ee
with $\eta$ a constant defined as in below (\ref{defPhi}) and $\zeta$ a quantised one-form. Therefore we have that the second term in the linear term (\ref{cristo}) is proportional to 
\be
\xi d\zeta-\varphi \mathcal L_X \tilde B\, =\, (\xi - \eta \varphi)d\zeta - \varphi\, P[\iota_X H] 
\ee
Since the D-brane equations of motion set $\zeta$ to be harmonic the first term cancels, while the second one can be seen as a potential generated for the D6-brane position in the presence of an $H$-flux. Similarly the first term in (\ref{cristo}) can be associated to a potential for $\varphi$ generated by $P[\iota_X dJ]$. Notice that in type IIA vacua $H + i dJ$ is related to the presence of a non-vanishig $W_0$. Hence we may interpret these terms as the piece of the scalar potential that comes from the piece $K_\Phi W_0$ of the covariant derivative. As explained in the main text and in appendix \ref{ap:4dsugra}, in the scenarios considered in this paper $W_0$ is taken very small or vanishing and therefore these potential terms for $\varphi$ can be neglected as compared to the scalar potential generated by (\ref{Winf}). 

One may now apply this reasoning to the general setup that we are interested in. Taking $\zeta$ harmonic will remove the Wilson line $\xi$ dependence from $Q_1$ and $Q_2$, and taking $\iota_X J_c|_{\Pi_3} = (\eta +i) \zeta$ will simplify the position $\varphi$ dependence leaving terms that depend on $H + i dJ$ and its derivatives. Finally, since those terms are proportional to $W^0_{\rm flux}$ neglecting them is basically the same approximation as the one done in section \ref{s:scenarios} and appendix \ref{ap:4dsugra} when neglecting terms in the scalar potential of order $|W_{\rm mod}^0|$. In this approximation we should take $\tilde Q = 1$ and that is why the DBI potential of section \ref{s:lifting} matches the B-field dependence of the 4d supergravity potential in the small field limit. It would be quite interesting to generalise this matching of potentials for vacua in which $W^0_{\rm flux}$ is non-negligible.

Notice that in the above reasoning we have made the crude estimate that the internal wavefunction for the Wilson line and the position modulus are the same as in the BPS case. In principle one should solve the equations of motion that determine these wavefunctions for an arbitrary background and then plug them in the expression for $\tilde Q$. We however do not expect that their profile for these wavefunctions change when restricted to the inflationary trajectory (\ref{infTB}) in which $\Pi_3$ is a special Lagrangian three-cycle. In general they should not change much whenever the Kaluza-Klein modes of the D6-brane are sufficiently massive. It would be however interesting to obtain explicit expressions for such lowest KK mode wavefunctions either from the microscopic viewpoint or in terms of a 4d backreaction problem.

\section{4d supergravity analysis}
\label{ap:4dsugra}

As stressed in the main text, at small field values the inflationary potential can be described as a 4d F-term supergravity scalar potential containing all the scalars of the compactification. This allows to understand the interplay of the inflationary sector with all the other massive scalars of the compactification, and to see to what extent both sectors are decoupled. 

The purpose of this appendix is to analyse the 4d supergravity potential of the type IIA compactifications discussed in the main text and to obtain an effective potential for the inflaton sector from it, applying the philosophy of section \ref{s:scenarios} to both of the scenarios described there. We will then use this result to analyse the stability of the inflationary trajectory against giving a vev to those scalars of the inflationary sector which are not the inflaton. As we will see near the trajectory one can show that these other scalars are more massive than the inflaton, at least in the small field regime where the supergravity approximation is valid. While in general inflation takes place outside this regime, we take the supergravity result as a good indicator on whether the inflationary trajectory is stable after Planck suppressed corrections have been taken into account. This intuition is partially tested in section \ref{s:saxion}, where it is indeed found that the supergravity stability bounds are very mildly corrected in the DBI potential.

\subsection{Type IIA scalar potential and moduli fixing}

Let us consider the 4d supergravity scalar potential 
\be
V\, =\, e^K \left(K^{\a\bar{\b}}D_\a W D_{\bar{\b}} \bar{W} - 3|W|^2  \right) \quad \quad \quad \a, \b \, =\, N^K, T^a, \Phi
\label{ap:pot}
\ee
where $W$ is given by (\ref{Wtot}) and the K\"ahler potential is $K = K_K + K_Q$, with the first piece given by (\ref{KK}) and the second by either (\ref{KQPhi}) or (\ref{KQPhi2}). As discussed in the main text, we are interested in a superpotential of the form 
\be
W\, =\, W_{\rm mod}  + W_{\rm inf} 
\label{ap:Wtot}
\ee
where $W_{\rm inf}$ is given by (\ref{Winf}) and depends on a particular linear combination $T$ of K\"ahler moduli, while $W_{\rm mod}$ is given by (\ref{Wmod}). For simplicity we will consider the case where the latter contains no linear terms in $\Phi$ or $T$, and so it can be written as
\be
W_{\rm mod} \, =\, W_1 + W_2 T^2 + W_3 \Phi^2 + \dots
\label{ap:Wmod}
\ee
where $W_i$, $i=1,2,3$ are such that $\p_TW_i = \p_\Phi W_i \equiv 0$, and the dots contain terms with higher powers on $\Phi$ and $T$. Finally, let us apply the assumption of section \ref{s:scenarios} and assume that $K_K$ only depends on $T$ via $(\im\, T)^2$. Then it is easy to see that the F-terms $D_T W$ and $D_\Phi W$ vanish at the point $\Phi = T =0$.  

In the following we would like to evaluate the scalar potential (\ref{ap:pot}) dependence on $(\Phi, T)$ around the point $\Phi = T =0$ and at point in closed string moduli space selected by $W_{\rm mod}$ and the K\"ahler potential $K = K_K + K_Q$. For simplicity we will choose an scenario where all their F-term vanish, namely we take $N^K, T^a$ at a value such that they solve
\be
[D_{N^K} W_{\rm mod}]_{\Phi=0}\ =\, [D_{T^a} W_{\rm mod}]_{\Phi=0}\, =\, 0
\label{Fterm0}
\ee
assuming that all closed string moduli are fixed by these conditions, except perhaps the axionic component of $T$. Following the discussion in the main text, these set of equations can be interpreted as the conditions for a 4d supersymmetric vacuum in the absence of the D6-brane generating $W_{\rm inf}$.\footnote{Because the minimum of the potential is supersymmetric and hence a solution to the equations of motion the caveats raised in \cite{Andriot:2015aza} would not apply to this model. It would be interesting to extend our analysis to 4d type IIA vacua where the closed string F-terms do not vanish, like for instance the scenario in \cite{Palti:2008mg}.}  As in the main text we label by $W_{\rm mod}^0$ the value of $W_{\rm mod}$ at the point selected by (\ref{Fterm0}), noticing that in order to connect with the framework in \cite{Kallosh:2010xz} we need to consider $|W_{\rm mod}^0|$ very small. 

To proceed and analyse the scalar potential dependence on $\Phi$, $T$ around this point let us first split (\ref{ap:pot}) as $V = V_Q + V_K - 3e^K|W|^2$, where
\bea
\label{VQ}
V_Q & = & e^K \left(K^{\a\bar{\b}}D_\a W D_{\bar{\b}} \bar{W} \right) \quad \quad \a, \b = N^{{K}}, \Phi \\
 V_K & = & e^K \left(K^{T^a\bar{T}^b}D_{T^a} W D_{\bar{T}^b} \bar{W} \right) 
\label{VK}
\eea

To evaluate (\ref{VQ}) we consider the F-terms $F_{N_K}$ around $\Phi = T =0$ and up to first order in such fields. Namely we have
\be
D_{N^K} W \, =\, K_{N^K} W_{\rm inf} + D_{N^K} W_{\rm mod}
\ee
where
\be
D_{N^K} W_{\rm mod}\, =\, \p_{N^K} W_1 + K_{N^K}^{\Phi=0} W_1 + \dots
\label{DNKexp}
\ee
where we have expanded up to linear order in $\Phi$, $\bar\Phi$ and $T$. Due to (\ref{Fterm0}) the rhs of (\ref{DNKexp}) vanishes at this order of the expansion, and we can simply take $D_{N^K} W  = K_{N^K} W_{\rm inf}$. Similarly, for the F-term $F_\Phi$ we find
\be
D_\Phi W\, =\, \p_{\Phi}(W_{\rm inf} + W_3^0 \Phi^2) +  K_{\Phi} (W_{\rm inf} + W_{\rm mod}^0) + \dots
\ee
where $W_3^0$ is the value of $W_3$ at the point where closed string moduli are stabilised.

Plugging this into (\ref{VQ}) and using the identities
\bea
\label{id1}
K^{\Phi\bar{\Phi}} K_{\bar{\Phi}} + \sum_{L} K^{\Phi\bar{N}^{L}} K_{\bar{N}^{L}} & = & 0 \\
\sum_{\a,\b = N^K, \Phi} K_\a K^{\a\bar{\b}} K_{\bar{\b}} & = & 4 
\label{id2}
\eea
we are able to express $V_Q$ as a sum of two squares
\be
V_Q\, =\, e^K \left(K^{\Phi\bar\Phi} \left| \p_\Phi W_{\rm inf}  + 2\Phi W_3^0 + K_\Phi W_{\rm mod}^{0} \right|^2 + 4 \left| W_{\rm inf} \right|^2 \right) 
\label{VQfin} 
\ee

Identities (\ref{id1}) and (\ref{id2}) can be checked by direct computation, and they apply to both versions (\ref{KQPhi}) and (\ref{KQPhi2}) of the K\"ahler potential. They can be understood from the fact that adding $\Phi$ to the K\"ahler potential (\ref{KQ}) in either way can be seen as a change of coordinates in the complex structure moduli space. Indeed, on the one hand and as pointed out in \cite{kt11}, the usual type IIA no-scale condition should also hold in this new coordinate system, and in our setup such  condition translates into the identity (\ref{id2}). 

Eq.(\ref{id1}), on the other hand, can be seen as follows. One may rewrite the K\"ahler potential (\ref{KQPhi}) and (\ref{KQPhi2}) as $K_Q(Z)  =  -2\, {\rm log} \left(\frac{i}{4} \CF_{{K}{L}} \im\, Z^{{K}} \im\, Z^{{L}} \right)$, with $Z^K(N^K,\Phi,\bar\Phi)$. Then it is easy to see that the differential operator
\be
X_{\bar{\Phi}}\, =\, \p_{\bar{\Phi}} + 2i (\p_{\bar{\Phi}} \im\, Z^{K})  \p_{\bar{N}^{K}}
\ee
is such that $X_{\bar{\Phi}} K_Q =0$. Finally, by the results in subsection \ref{ap:metrics} one can check explicitly that
\be
\p_{\bar{\Phi}} Z^{K} \, =\, \frac{K^{\Phi\bar{N}^K}}{K^{\Phi\bar{\Phi}}}
\ee
and so (\ref{id1}) follows from applying $X_{\bar{\Phi}}$ on $K_Q$. 

One may now evaluate (\ref{VK})  by using the following F-terms
\bea
\label{ap:DTaW}
D_{T^\a} W & = & K_{T^\a} W_{\rm inf} + D_{T^\a} W_{\rm mod}\, =\, K_{T^\a} W_{\rm inf} + \dots \\ 
D_T W & = & \p_T (W_{\rm inf} + W_2^0 T^2) + K_T [W_{\rm mod}^0 + W_{\rm inf}]  + \dots
\label{ap:DTW}
\eea
where $T^\a$ are the K\"ahler moduli that $W_i$ depend on, and where $W_2^0$ is the value of $W_2$ at the point where closed string moduli are stabilised. Again we have expanded up to linear order in $T$, $\bar T$ and $\Phi$ and imposed (\ref{Fterm0}). Plugging these expressions into (\ref{VK}) and using the identities (\ref{identiT}) we find that
\bea\nonumber
V_{K} & = & e^K\left(K^{T\bar T} |\p_TW_{\rm inf} + 2T W_2^0 + K_T W_{\rm mod}^0|^2 + (2i\im \, T)^2 |\p_TW_{\rm inf}|^2 +  3\left| W_{\rm inf} \right|^2  \right) \\
& + & e^K\left(\sum_a K_{T^a} K^{T^a\bar T}  W_{\rm inf} (K_{\bar T} \overline{W}_{\rm mod}^0 + 2 \bar{T}\overline{W}_2^0)  + \text{c.c.}\right)
\label{VKfin}
\eea
Finally, adding (\ref{VQfin}) and (\ref{VKfin}) into $V = V_Q + V_K - 3e^K|W|^2$ we obtain
\bea\nonumber
V & = & e^K\left( K^{\Phi\bar\Phi} \left| \p_\Phi W_{\rm inf} + 2\Phi W_3^0 + K_\Phi W_{\rm mod}^{0} \right|^2 + K^{T\bar T} |\p_TW_{\rm inf} + 2TW_2^0+K_T W_{\rm mod}^0|^2  \right)   \\ \nonumber
 & + & e^K\left(4 (\re\, T)^2 |\p_TW_{\rm inf}|^2 + (4 i \im\, T K_{\bar T} - 6)\,  \re (W_{\rm inf} \overline{W}_{\rm mod}^0) + \re ( (8i \im\, T) \bar{T} W_{\rm inf} \overline{W}_{2}^0) \right) \\
 & - & 3 e^K |W_{\rm mod}^0|^2
\label{Vfin}
\eea
Notice that the first line of (\ref{Vfin}) contains the terms quadratic in $\Phi$ and $T$ and hence determines the mass matrix for these fields. The third line contains a constant term which  is nothing but the vacuum energy inherited from the closed string moduli stabilisation process. Finally, the second line contains various terms with quartic dependence dependence on $\Phi$ and $T$. While at the level of approximation which we have taken one may in principle neglect these terms, they contain a non-trivial dependence on the inflaton candidates $\re \, T$ and $\re\, \Phi$, so they may become relevant in each of the two scenarios discussed in section \ref{s:scenarios}. In the following we analyse both scenarios and adapt the computation that led to the expression (\ref{Vfin}) for each of them.  

\subsubsection*{Inflating with the B-field}

Let us fist consider the scenario of section \ref{Bfield}. There, on top of the assumptions already taken it was assumed that $W_{\rm mod}$ does not depend on the K\"ahler modulus $T$, so that the B-field direction $\re\, T$ is a flat direction of the scalar potential if we switch off $W_{\rm inf}$.\footnote{Alternatively, one may consider the case where $W_2^0$ is very small, so that the mass contribution to $\re\, T$ from $W_{\rm mod}$ is extremely small. This case, however, is quite analogous to the one analysed in \cite{Hebecker:2014kva} and we would expect that it suffers from the problems of fine-tuning and backreaction there discussed.} Imposing such extra condition on (\ref{ap:Wmod}) implies, in particular, that $W_2 \equiv 0$, and applying it to the computation above gives
\bea\nonumber
V & = & e^K\left( K^{\Phi\bar\Phi} \left| \p_\Phi W_{\rm inf} + 2\Phi W_3^0 + K_\Phi W_{\rm mod}^{0} \right|^2 + K^{T\bar T} |\p_TW_{\rm inf} +K_T W_{\rm mod}^0|^2  + 4 (\re\, T)^2 |\p_TW_{\rm inf}|^2 \right)   \\ 
 & + & e^K\left( (4 i \im\, T K_{\bar T} - 6)\,  \re (W_{\rm inf} \overline{W}_{\rm mod}^0) - 3  |W_{\rm mod}^0|^2 \right)
\eea
One can check that this expression for the potential is exact in the inflaton candidate $\re\, T$, while it is quadratic in the fields $\Phi$, $\bar\Phi$ and $\im\, T$. If we now take $|W_{\rm mod}^0|$ very small in order to connect with the setup of \cite{Kallosh:2010xz} the second line can be neglected, and one finds
\be
V \, = \, e^K\left( K^{\Phi\bar\Phi} \left| \p_\Phi W_{\rm inf} + 2\Phi W_3^0\right|^2 + (K^{T\bar T}  + 4 (\re\, T)^2) |\p_TW_{\rm inf}|^2 \right)  + \CO(W_{\rm mod}^0)
\label{VBfinapprox}
\ee
Finally, if we impose the condition $\p_\Phi W_{\rm mod} =0$ then $W_3 \equiv 0$ and we recover the result in \cite{Escobar:2015fda}.

\subsubsection*{Inflating with the Wilson line}

We now consider the scenario of section \ref{WLine}, in which the inflaton candidate is the D6-brane Wilson line $\re\, \Phi$. In this case we require that $K_Q$ is given by (\ref{KQPhi2}) and that $W_{\rm mod}$ does not depend on $\Phi$, which implies that $W_3\equiv 0$. In this case we obtain that the scalar potential is
\bea\nonumber
V & = & e^K\left( K^{\Phi\bar\Phi} \left| \p_\Phi W_{\rm inf} + K_\Phi W_{\rm mod}^{0} \right|^2 + K^{T\bar T} |\p_TW_{\rm inf} + 2TW_2^0+K_T W_{\rm mod}^0|^2  + 4 |a|^2 (\re\, T)^2 (\re\, \Phi)^2 \right)   \\ 
 & - & e^K\left(6\,  \re (W_{\rm inf} \overline{W}_{\rm mod}^0) + 3 |W_{\rm mod}^0|^2\right) 
\eea
where we have neglected terms of cubic order in $T$, $\bar{T}$ and $\im\, \Phi$. Again, one can check that otherwise the above expression is exact in $\re\, \Phi$, and therefore it can be used along the inflationary trajectory up to the point where the supergravity approximation is not trustable. Finally, taking the limit of very small $|W_{\rm mod}^0|$ we obtain
\be
V  =  e^K\left( K^{\Phi\bar\Phi} \left| \p_\Phi W_{\rm inf} \right|^2 + K^{T\bar T} |\p_TW_{\rm inf} + 2TW_2^0|^2  + 4 |a|^2 (\re\, T)^2 (\re\, \Phi)^2 \right) + \CO(W_{\rm mod}^0)
\label{VWfinapprox}
\ee

\subsection{Effective potentials and stability bounds}

Given the above scalar potentials, one must consider the stability of the inflationary trajectory for each of them. That is, since out the two complex fields $\Phi$ and $T$ we have selected one real scalar as the inflaton candidate, we must insure that all the other three real directions remain non-tachyonic during inflation. Finally, in order to describe our system as a model of single field inflation these three scalars must have a mass higher than the Hubble scale, since otherwise they cannot be decoupled from the inflationary dynamics. 

This sort of analysis was carried in \cite{Kallosh:2010xz} for a rather general class of supergravity chaotic inflation models with a stabiliser field. The main results were then encoded in two stability bounds expressed in terms of a normalised K\"ahler potential. For the models analysed in \cite{Kallosh:2010xz}, if such inequality bounds are satisfied then the three scalar fields beyond the inflaton are massive enough to be decoupled during inflation. The case of interest in this paper is different from the models in  \cite{Kallosh:2010xz}, in the sense that  the effective scalar potential is derived after a process of moduli stabilisation that has been analysed in the previous section. As a result extra terms appear in the potential as compared to the potentials in \cite{Kallosh:2010xz}, and so the whole analysis must be reconsidered. In the following we will perform such analysis for the scalar potential derived above, both for the case where the inflaton is a B-field or a Wilson line axion. In both scenarios we will find that the extra terms obtained from the process of moduli fixing relax the stability bounds found in \cite{Kallosh:2010xz}, making them easier to satisfy. 

\subsubsection*{Inflating with the B-field}

In this scenario the inflationary trajectory is given by
\be
\text{Traj} = \left\lbrace \text{Re{\it T}} \neq 0  \text{ ,  } \text{Im{\it T}} = 0 \text{ ,  }  \Phi = 0  \right\rbrace
\ee
and the scalar potential is (\ref{VBfinapprox}). Because $W_3$ in (\ref{ap:Wmod}) arises from either worldsheet or D-brane instanton effects it will be naturally suppressed with respect to other terms in the superpotential, and so we may approximate $W_3^0 \simeq 0$. The effective potential then reduces to
\be
V \, = \, |a|^2 e^K\left( K^{\Phi\bar\Phi}  \left| T \right|^2 + (K^{T\bar T}  + 4 (\re\, T)^2) |\Phi|^2 \right)  
\label{VBfinal}
\ee
and one can check that the trajectory is an extremum in the non-inflationary directions, namely
\be
\p_{\text{Im}\, T} V|_{\text{Traj}}\, =\, \p_{\Phi} V|_{\text{Traj}}\, =\, \p_{\bar\Phi} V|_{\text{Traj}} \, =\, 0\, .
\ee
A more constraining requirement arises from demanding that the masses of these three fields are beyond the Hubble scale. For the canonically normalised saxion partner of the inflaton we have that
\be
m_{\rm saxion}^{2}\mid_{\rm Traj}\, =\, \frac{1}{2K_{T\bar T}} \p_{\text{Im}T}^{2} V\mid_{\rm Traj} 
\label{ap:saxionmass}
\ee
and so
\bea\nonumber
m_{\rm saxion}^{2}\mid_{\rm Traj} &  = & |a|^2 e^K K^{\Phi\bar{\Phi}} \left(K_{T\bar T}^{-1}   + \left[2 + \frac{\p^2_{\text{Im}T}K^{\Phi\bar{\Phi}}}{2K_{T\bar T} K^{\Phi\bar{\Phi}}} \right]  (\re \, T)^2  \right)_{\text{Traj}}\\
& \simeq & 3H^2 \left(\eps + 2 + \frac{\p^2_{\text{Im}T}K^{\Phi\bar{\Phi}}}{2K_{T\bar T} K^{\Phi\bar{\Phi}}} \right)_{\text{Traj}}
\label{saxionm1}
\eea
where we have used our assumption that $K$ only depends on $T$ via $(\im\, T)^2$ which implies that
\be
K_{T\bar{T}} = -K_{TT} = -K_{\bar{T}\bar{T}}
\label{kahsym2}
\ee
and identified the cosmological parameters as
\be
3H^2\, \simeq \, |a|^2 e^K K^{\Phi\bar{\Phi}} (\re \, T)^2   \quad \quad \quad \quad \eps\, =\, \frac{1}{K_{T\bar{T}} (\re \, T)^2}
\ee
evaluated at the trajectory. Because the K\"ahler potential split as $K  = K_K(T^a) + K_Q(N^K, \Phi)$, $K^{\Phi\bar\Phi}$ does not depend on $\im T$ and so the last contribution to (\ref{saxionm1}) vanishes. Moreover, because during inflation $\eps \ll 1$ the first contribution can be neglected and so we arrive at
\be
m_{\rm saxion}^{2}\mid_{\rm Traj}\, \simeq\, 6 H^2
\label{saxionm1fin}
\ee
which satisfies the criteria drawn in \cite{Kallosh:2010xz}. 

Regarding the open string field that here plays the role of stabiliser we have that the normalised fields are $s_1$ and $s_2$ where $s_1 + i s_2 = \sqrt{2 K_{\Phi\bar\Phi}} \Phi$ and so
\be
m_{s_1}^{2}\mid_{\rm Traj}\, =\, \frac{1}{2K_{\Phi\bar \Phi}} \p_{\text{Re}\Phi}^{2} V\mid_{\rm Traj} \quad \quad m_{s_2}^{2}\mid_{\rm Traj}\, =\, \frac{1}{2K_{\Phi\bar \Phi}} \p_{\text{Im}\Phi}^{2} V\mid_{\rm Traj}
\ee
The precise expressions for these two masses depends on the expression for the K\"ahler potential piece $K_Q$, and in particular on whether we should consider (\ref{KQPhi}) or (\ref{KQPhi2}). For simplicity we here consider the first choice (\ref{KQPhi}), for which we have that both masses are equal to
\bea\nonumber
m_{\rm stab}^{2}\mid_{\rm Traj} & =& |a|^2 e^K K^{\Phi\bar{\Phi}} \left(K^{T\bar T}   + \left[4 + 1 - \oh (K^{\Phi\bar\Phi})^2 K_{\Phi\bar\Phi\Phi\bar\Phi} \right]  (\re \, T)^2  \right)_{\text{Traj}}\\
& \simeq & 3H^2 \left( K^{T\bar T} K_{T\bar T}\, \eps + 5 - \oh (K^{\Phi\bar\Phi})^2 K_{\Phi\bar\Phi\Phi\bar\Phi} \right)_{\text{Traj}} 
\label{stabm1}
\eea
where we have used that at $\Phi=\bar\Phi =0$
\be
K^{\Phi\bar\Phi} \, =\, (K_{\Phi\bar\Phi})^{-1} \quad \quad {\rm and} \quad \quad \quad  \p_{\Phi}\p_{\bar\Phi}  K^{\Phi\bar\Phi} = -\oh (K^{\Phi\bar\Phi})^2 K_{\Phi\bar\Phi\Phi\bar\Phi}
\ee
as follows from the results of appendix \ref{ap:metrics}. On can also check that, because $K_K$ only depends on $T$ via $(\im\, T)^2$, $K^{T\bar T} K_{T\bar T} = 1$ at $\im\, T =0$ and so the first term in (\ref{stabm1}) can be neglected.  We are then left with
\be
m_{\rm stab}^{2}\mid_{\rm Traj}\, \simeq \, 3H^2 \left(5 - \oh (K^{\Phi\bar\Phi})^2 K_{\Phi\bar\Phi\Phi\bar\Phi} \right)_{\text{Traj}} \, .
\label{stabm1fin}
\ee
Compared to the result in \cite{Kallosh:2010xz} there is an extra contribution of $15 H^2$ that pushes the stabiliser mass above the Hubble scale. The second contribution is similar to the one found in \cite{Kallosh:2010xz}, and it may be positive or negative depending on the parameters of the compactification. 

Indeed, in order to evaluate this second term let us first rewrite (\ref{KQPhi}) as
\be
K_{Q}  =  -  2 {\rm log}\left(\CF^{\, 0}\right) - 2 {\rm log}\left(1 + \frac{i}{2} \left(\Phi \bar{\Phi}\right) \frac{\partial_{N^{K}}\CF^{\, 0} Q^{K}}{\CF^{\,0}} - \frac{1}{16} \left(\Phi \bar{\Phi}\right)^2 \frac{\partial_{N^{K}}\partial_{N^{L}}\CF^{\, 0} Q^{K}Q^{L}}{\CF^{\, 0}}\right)
\ee
where we have defined
\be
\CF^{\, 0} \, =\, \frac{1}{16i} \CF_{{K}{L}} \left[N^{{K}} - \bar{N}^{{K}}  \right] \left[N^{{L}} - \bar{N}^{{L}}  \right] 
\ee
We may now expand the second logarithm around $x = \Phi \bar{\Phi}$ as
\be
-2 {\rm log}\left(1 +Ax + Bx^2\right) \simeq -2Ax + \left(A^2 - 2B\right)x^2 + \mathcal{O}\left(x^3\right)
\label{expansion}
\ee
obtaining that the coefficient for $\Phi\bar\Phi$ is given by
\be
K_{\Phi\bar\Phi}|_{\Phi=0}\, =\, -i \frac{\partial_{N^{K}}\CF^0 Q^{K}}{\CF^0}\, =\, -\oh \frac{\CF_{KL}\text{Im}N^{L}Q^{K} }{\CF_{KL}\text{Im}N^{K}\text{Im}N^{L}}\, =\, (K^{\Phi\bar\Phi}|_{\Phi=0})^{-1}
\ee
in agreement with eq.(\ref{KPP}). From the coefficient of $(\Phi\bar\Phi)^2$ one obtains that
\be
-  \oh \left.(K^{\Phi\bar\Phi})^2 K_{\Phi\bar\Phi\Phi\bar\Phi} \right|_{\text{Traj}}\, =\, \frac{1}{4} \left(\frac{{\CF}_{KL}{Q}^{K}{Q}^{L}{\CF}_{RS}\text{Im}N^{R}\text{Im}N^{S}}{2\left({\CF}_{KL}\text{Im}N^{L}{Q}^{K}\right)^2} -1 \right)
\label{quarticphi}
\ee
the first term depending on where the complex structure fields are stabilised. Generically, one would expect that this term is an order one positive number, obtaining that the stabiliser field mass at the trajectory is above the Hubble scale. It would be however interesting to evaluate the quantity (\ref{quarticphi}) for explicit models with concrete mechanisms and values for complex structure moduli stabilisation. 

\subsubsection*{Inflating with the Wilson line}

In this case the inflationary trajectory is given by
\be
\text{Traj} = \left\lbrace \re\, \Phi \neq 0  \text{ ,  } \im\, \Phi = 0 \text{ ,  }  T = 0  \right\rbrace
\ee
and the scalar potential is (\ref{VWfinapprox}). In this case $W_2$ can arise from a flux superpotential and it may be as large as any other term, but in order to simplify the discussion we will assume that $W_2^0 = 0$, leaving the more general case for future work. The effective potential then reads
\be
V \, = \, |a|^2 e^K\left( K^{\Phi\bar\Phi}  \left| T \right|^2 + K^{T\bar T} |\Phi|^2 + 4 (\re\, T)^2  (\re\, \Phi)^2 \right)  
\label{VWLfinal}
\ee
and one can easily check that 
\be
\p_{\text{Im}\, \Phi} V|_{\text{Traj}}\, =\, \p_{T} V|_{\text{Traj}}\, =\, \p_{\bar T} V|_{\text{Traj}} \, =\, 0\, .
\ee
The first stability bound is now expressed in terms of
\be
m_{\rm saxion}^{2}\mid_{\rm Traj} \, = \, \frac{1}{2K_{\Phi\bar\Phi}} \p_{\text{Im}\Phi}^{2} V\mid_{\rm Traj} \, \simeq \, 3H^2 \left( \eps +2 \right) \, \simeq \, 6H^2
\ee
where we have again used that $K$ splits as split as $K  = K_K(T^a) + K_Q(N^K, \Phi)$, and now
\be
3H^2\, \simeq \, |a|^2 e^K K^{T\bar{T}} (\re \, \Phi)^2   \quad \quad \quad \quad \eps\, =\, \frac{1}{K_{\Phi\bar{\Phi}} (\re \, \Phi)^2}
\ee
The stability bound for the stabiliser field is more involved, and it turns to be different for the real and imaginary parts. Now defining $s_1 + i s_2 = \sqrt{2 K_{T\bar T}} T$ we have that
\be
m_{s_1}^{2}\mid_{\rm Traj}\, =\, \frac{1}{2K_{T \bar T}} \p_{\text{Re}T}^{2} V\mid_{\rm Traj} \quad \quad m_{s_2}^{2}\mid_{\rm Traj}\, =\, \frac{1}{2K_{T \bar T}} \p_{\text{Im} T}^{2} V\mid_{\rm Traj}
\ee
and so
\be
m_{s_1}^{2}\mid_{\rm Traj}\, =\, |a|^2 e^K (K_{T\bar T})^{-1} \left( K^{\Phi\bar\Phi} + 4  (\re\, \Phi)^2  \right)_{\rm Traj} \, \simeq \, 3H^2 \left( \eps + 4\right)\, \simeq\, 12 H^2
\ee
where we have used that $K^{T\bar T}$ only depends on $\im T$. Similarly
\bea\nonumber
m_{s_2}^{2}\mid_{\rm Traj} &  = &|a|^2 e^K K^{T\bar T} \left( K^{\Phi\bar\Phi} + \left( 2 + \oh \p^2_{\text{Im}T}K^{T\bar{T}} \right) (\re \, \Phi)^2  \right)_{\rm Traj}\\
& \simeq & 3H^2 \left( \eps + 2 \left[ 1 +  \p_T\p_{\bar{T}}K^{T\bar{T}} \right]_{\rm Traj} \right)\, \simeq\, 6H^2 \left(  1 +  \p_T\p_{\bar{T}}K^{T\bar{T}}  \right)_{\rm Traj}
\eea
and so in the second case the mass will depend on the stabilisation details for the K\"ahler moduli. 

\subsection{K\"ahler metrics}{\label{ap:metrics}

The 4d K\"ahler metric in our setup is given by 
\be
{\bf K}\ =\, 
\left(
\begin{array}{cc}
{\bf K_K} \\ & {\bf K_Q}
\end{array}
\right)
\ee
where with a slight abuse of notation we have defined the matrices
\bea
\label{metricKK}
({\bf K_K})_{a\bar{b}} & \equiv &  \p_{T^a}\p_{\bar{T}^b} K_K = K_{a\bar{b}} \\
({\bf K_Q})_{\a\bar\b} & \equiv &\p_\a\p_{\bar\b} K_Q\, =\, K_{\a\bar\b} \quad \quad \a, \b = N^{{K}}, \Phi
\label{metricKQ}
\eea
where in the rhs of (\ref{metricKK}) $K_K$ is given by (\ref{KK}) and in the rhs of (\ref{metricKQ}) $K_Q$ is given by either (\ref{KQPhi}) or (\ref{KQPhi2}), and $K = K_K + K_Q$.

In order to find the inverse of the matrix ${\bf K_Q}$ notice that it is of the form 
\be
{\bf K_Q}\, =\, 
\left(
\begin{array}{cc}
A & -AB \\ -B^\dag A & B^\dag A B + C
\end{array}
\right)
\, =\,
\left(
\begin{array}{cc}
\Id & 0 \\ -B^\dag  & 1
\end{array}
\right)
\left(
\begin{array}{cc}
A & 0 \\ 0 &  C
\end{array}
\right)
\left(
\begin{array}{cc}
\Id & -B \\ 0 &  1
\end{array}
\right)
\label{KQmat}
\ee
where 
\be
A_{KL}\, =\, K_{N^K\bar{N}^L} \quad\quad B^L \, =\, \p_{\bar\Phi} Z^L \quad \quad C\, =\, \frac{i}{4}K_{N^K}Q^K
\ee
and where as above we have defined $Z^L$ by writing $K_Q(Z)  =  -2\, {\rm log} \left(\frac{i}{4} \CF_{{K}{L}} \im\, Z^{{K}} \im\, Z^{{L}} \right)$. The inverse of (\ref{KQmat}) is given by
\be
{\bf K_Q}^{-1} \, =\, 
\left(
\begin{array}{cc}
A^{KL} + C^{-1} B^KB^{\dag\, L} & C^{-1} B^L \\  C^{-1} B^{\dag\, K} &  C^{-1}
\end{array}
\right)
\label{KQinv}
\ee
with $A^{KL}$ the inverse of $A_{KL}$. From here we obtain that
\be
\frac{K^{\Phi\bar{N}^K}}{K^{\Phi\bar{\Phi}}}\, =\, \p_{\bar\Phi} Z^K\quad \quad K^{\Phi\bar\Phi} \, =\, \left[\frac{i}{4}K_{N^K}Q^K\right]^{-1}
\label{Kinv}
\ee

To analyse the inverse of ${\bf K_K}$ it is useful to define the following quantities
\be
\CK_{ab}=\CK_{abc}v^c \qquad  \CK_{a}=\CK_{abc}v^bv^c\qquad  \CK=\CK_{abc}v^av^bv^c.
\label{defsK}
\ee
with $v^a=e^{-\phi/2}\im\, T^a$. We then have the following derivatives of \eqref{KK}
\be
K_a=\frac{3i}{2\CK}e^{-\phi/2}\CK_a \qquad \quad K_{a\bar b}=-\frac{3}{2\CK^2}e^{-\phi}\left (\CK\CK_{ab}-\frac{3}{2}\CK_a\CK_b\right ).
\label{derivKK}
\ee
and so the inverse metric is given by
\be
K^{a\bar b}=-\frac{2}{3}e^{\phi}\CK\CK^{ab}+2e^{\phi}v^av^b
\ee
where $\CK^{ab}$ is the inverse of $\CK_{ab}$ which implies that
\be
\CK^{ab}\CK_b=v^a.
\ee
One can check that indeed $K^{a\bar b}K_{c\bar b}=\delta^a_c$ and $K^{a\bar b}K_{a\bar c}=\delta^{\bar a}_{\bar c}$. Finally we also have that
\be
K_aK^{a\bar b}K_{\bar b}=3,\qquad \quad K_a K^{a\bar b}=2ie^{-\phi/2}v^b.
\label{identiT}
\ee

\section{A simple background for the Wilson line scenario}
\label{WLtoy}

In the following we will consider a simple type IIA compactification that can be used as a toy model for implementing the scenario is the Wilson line. More precisely, we will consider the class of type IIA flux compactifications studied in \cite{DeWolfe:2005uu} and see under which conditions one can have a closed string background with the properties described in section \ref{WLine}. As a second step one should add a D6-brane wrapping a three-cycle with the required topology described in section \ref{s:lifting}, a task that we leave for future analysis.

For simplicity let us consider a type IIA compactification with two K\"ahler moduli, which we dub $T_1$ and $T_2$. We may then define the linear combinations
\be
T_{+} \, = \, \frac{1}{2}\left(T_{1} + T_{2}\right) \quad \quad {\rm and} \quad \quad T_{-} \, = \, \frac{1}{2}\left(T_{1} - T_{2}\right)
\ee
and identify $T_-$ with the combination of K\"ahler moduli (\ref{defT}) that will appear in the the bilinear superpotential $W_{\rm inf}$ when we add the D6-brane, and which we have dubbed $T$ in the main text. From this example it is easy to see that $T_-=0$ does not imply that any volume of the of the compactification vanishes, but rather that two compactification volumes are related. 

One of the requirements for both scenarios of section \ref{s:scenarios} is that the K\"ahler potential of the compactification only depends on $T$ through $(\im\, T)^2$. In the case at hand and taking $K = K_K + K_Q$ we see that this is easily achievable by imposing the following relations for the triple intersection numbers
\be
\CK_{111}\, =\, \CK_{222} \quad \quad {\rm and} \quad \quad \CK_{122}\, =\, \CK_{211}
\ee
From here we obtain
\be
K_K\, =\, -{\rm log} \left(\frac{i}{6} \CK_{+++} (T_+ - \bar{T}_+)^3 + \frac{i}{2} \CK_{+--} (T_{-} - \bar{T}_{-})^2 (T_{+} - \bar{T}_{+}) \right) 
\label{KKap}
\ee
where we have defined
\be
\CK_{+++}\, =\, 2(\CK_{111} + 3\CK_{112}) \quad \quad {\rm and} \quad \quad \CK_{+--}\, =\, 2(\CK_{111} - \CK_{112})\, .
\ee

An extra requirement of the Wilson line scenario is that there appear no linear terms in $T=T_-$ in $W_{\rm mod}$. To evaluate this condition let us consider the class of type IIA compactifications considered in \cite{DeWolfe:2005uu}, in which
\be
W_{\rm mod} \, =\, W_{\rm flux}\, =\, W_K + W_Q
\ee
where $W_K$ is given by (\ref{WK}) and 
\be
W_{Q} = \int \Omega_{c} \wedge H_{3} = -N^{K} p_{{K}}\, =\,-\left(\xi^{{{K}}} + i \text{Re}\left(e^{-\phi}C Z^{{K}}\right)\right) p_{{{K}}}  =\,- \left(\xi^{{{K}}} + i l^{{{K}}}\right) p_{{{K}}}
\ee
with the moduli $N^K$ defined as in (\ref{defNK}). In this case obtaining a superpotential with no linear term in $T_-$ is achievable by imposing the following relations among RR background fluxes
\be
e_1\, =\, e_2 \, =\, e  \quad \quad {\rm and} \quad \quad m_1\, =\, m_2 \, =\, m 
\ee
from which we obtain that
\bea\nonumber
W_K & = & e_0 + 2 e T_+ +   {\frac{m}{2}} \CK_{+++} T_+^2 +  {\frac{m}{2}} \CK_{+--} T_-^2 - \frac{1}{6} m_0 \left( \CK_{+++} T_+^3 + 3 \CK_{+--} T_+T_-^2 \right) \\
&= & W_1(T_+) + \frac{1}{2}\CK_{+--} \left(m -  m_0   T_+ \right) T_-^2
\eea
as required in the main text.

\subsubsection*{Moduli stabilisation}

Let us now compute the point in moduli space in which the closed string moduli are stabilised with vanishing F-terms. That is, we impose the conditions
\be
D_{T^a}W_{\rm mod}\, =\, D_{N^K}W_{\rm mod}\, =\, 0
\ee
with the superpotential above and the K\"ahler potential $K = K_K + K_Q=$(\ref{KKap})+(\ref{KQ}). 
Following the general discussion in \cite{DeWolfe:2005uu} we first consider the stabilisation of the complex structure moduli, whose F-term is given by
\be
D_{N^{{{K}}}} W_{\text{mod}} = -p_{{{K}}} + K_{N^{{K}}}W_{\rm mod} {= p_{K} + 4e^{2D}\mathcal{F}_{KL}l^{L}W_{\text{mod}}} = 0
\label{FTermcxstr}
\ee
Note that $\mathcal{F}_{{K}{L}}$ is pure imaginary by definition \cite{kt11}. Analysing its imaginary part we arrive to
\be
\text{Re}\, W_{\text{mod}} = 0 \quad \Rightarrow\quad  -p_{{K}}\xi^{{K}} + \text{Re}\, W_{K} = 0 
\label{realFtermcxstr}
\ee
which implies that only a linear combination of RR three-form axions will be stabilised by the fluxes. Notice however that when we include D6-branes some other linear combinations will  be eaten by open string gauge bosons and become massive via St\"uckelberg mechanism, and therefore they should not appear in the superpotential \cite{Camara:2005dc}. Hence the lack of stabilisation of some of these axions should not be seen as a flaw of the model but rather as a necessary condition to introduce D6-branes, which is important for our purposes.
The real part of (\ref{FTermcxstr}) will give us
\be
p_{K} + 4i e^{2D}{\cal F}_{{K}{L}}l^{{L}}\text{Im}\, W_{\text{mod}}  = 0
\ee
Where we have used that $e^{-2D} = 2il^{{K}}{\cal F}_{{K}{L}}l^{{L}}$. Note that $\text{Im } W =0$ implies zero $H_{3}$ flux. For $\text{Im}W \ne 0$ we see that for every $p_{K_{i}}$ different from zero and rearranging the former expression we arrive to
\be
\frac{ip_{ {K_{i}}}}{{\cal F}_{ {K_{i}} {L}}Z^{ {L}}}e^{-K_{CS}/2} \, =\,  \frac{p_{ {K_{i}}}}{\text{Im}{\cal F}_{ {K_{i}}}}e^{-K_{CS}/2}\, :=\, Q_0
\label{4Ddilaton}
\ee
where we have used the the definition $l^{{K}}:=e^{-D}e^{\frac{1}{2}K_{CS}}Z^{{K}}$ and the relation $\CF_K = \CF_{ {K} {L}}Z^{L}$, note that $Q_0$ is a fixed quantity. The above system  of $h^{2,1}$ equations, generically, will stabilise all the complex structure saxions to a specific value. Finally using that $e^{D} = e^{\phi + \frac{1}{2}K_{K}} = \frac{e^{\frac{\phi}{4}}}{\sqrt{\frac{4}{3}\mathcal{K}}} $ we find that the dilaton is stabilised at
\be
e^{-\phi} = 4 \frac{e^{K_{K}/2} }{Q_{0}}\text{Im}\, W_{\text{mod}}
\ee
Regarding the F-terms for the K\"ahler moduli, first of all we will derive that the superpotential evaluated in the vacuum can be written only in terms of the K\"ahler moduli
	\be
	W_{\text{mod}} = -i \text{Im }W_{K}
	\ee
	this can be see taking into account the following: if we multiply (\ref{FTermcxstr}) per $l^{K}$ and sum over $K$, and using the definition of $D$ we arrive to
	\be
	-iW_{\text{mod}} = \frac{1}{2}\text{Im }W_{Q}
	\ee
Imposing the above relations for the complex structure moduli we find that the F-term equation for the K\"ahler moduli is given by
\be
D_{T_{a}}W_{K} - i K_{T_{a}} \text{Im}\, W_{K}=0
\label{F-termKah}
\ee
whose imaginary part fixes the B-field axions to
\be
\text{Re}\, T_{+} \, = \,  {m}/m_{0} \quad \text{and} \quad  \text{Re}\, T_{-} \, =\, 0
\label{stabaxions}
\ee
Moreover, one can see that the real part of the F-term for $T_-$ fixes $\im\, T_- = 0$ while that for $T_+$ imposes the relation
\be
 {20}e + \CK_{+++}\left(3 m_{0}\, \text{Im}\, T_{+}^2 +  {5} \frac{m^2}{m_0} \right) = 0
\ee
so the volume modulus is stabilised at
\be
\text{Im}\, T_{+} =  \frac{\sqrt{5}}{\sqrt{3}m_{0}}\sqrt{-\frac{ {4}e m_{0}}{\CK_{+++}} -   m^{2}}
\ee
which is positive as long as $e<0$ and $|e|m_0 >  {\frac{1}{4}}m^2 \CK_{+++}$.

\end{document}